\documentclass[
 preprint,
 preprintnumbers,
 amsmath,amssymb,
 aps,
 pre,
 longbibliography,
 superscriptaddress,
 10pt, twocolumn
]{revtex4-1}

\usepackage{graphicx}
\usepackage{dcolumn}
\usepackage{bm}
\usepackage{hyperref}
\usepackage{physics} 
\usepackage{color}
\begin{document}
\newcommand{\eqnname}{Eq.}
\newcommand{\secname}{Sec.}

\title{Rectification of energy and motion in non-equilibrium parity violating metamaterials}

\author{Zhenghan Liao}
\affiliation{Department of Chemistry, University of Chicago, Chicago, IL, 60637, USA}
\author{William T. M. Irvine}
\affiliation{Department of Physics, University of Chicago, Chicago, IL 60637, USA}
\affiliation{James Franck Institute, University of Chicago, Chicago, IL 60637, USA}
\affiliation{ Enrico Fermi Institute, University of Chicago, Chicago, IL, 60637, USA}
\author{Suriyanarayanan Vaikuntanathan}
\affiliation{Department of Chemistry, University of Chicago, Chicago, IL, 60637, USA}
\affiliation{James Franck Institute, University of Chicago, Chicago, IL 60637, USA}

\begin{abstract}
Uncovering new mechanisms for rectification of stochastic fluctuations has been a longstanding problem in non-equilibrium statistical mechanics. Here, using a model parity violating metamaterial that is allowed to interact with a bath of active energy consuming particles, we uncover new mechanisms for rectification of energy and motion. Our model active metamaterial can generate energy flows through an object in the absence of any temperature gradient. The nonreciprocal microscopic fluctuations responsible for generating the energy flows can further be used to power locomotion in, or exert forces on, a viscous fluid. Taken together, our analytical and numerical results elucidate how the geometry and inter-particle interactions of the parity violating material can couple with the non-equilibrium fluctuations of an active bath and enable rectification of energy and motion.
\end{abstract}

\maketitle

\section{Introduction} \label{sec:intro}

Identifying mechanisms that rectify stochastic fluctuations is a longstanding problem in non-equilibrium statistical mechanics ~\cite{Seifert2012StochasticThermodynamics,Coskun2011GreatExpectations}. The Feynman Ratchet and pawl model, and its associated generalizations, have elucidated how systems can rectify stochastic fluctuations and act as microscopic engines that perform work and exert forces~\cite{Jarzynski1999FeynmanRatchet}. 
Indeed such models have provided a framework to understand how biological molecular motors can convert the energy derived from the hydrolysis of energy rich molecules into mechanical work~\cite{Mogilner1996CellMotility,Chernyak2008PumpingRestriction,Rahav2008DirectedFlow,Sinitsyn2007UniversalGeometric}. 
While these advances provide powerful insights into how single-body systems can rectify fluctuations,  much less is understood about rectification via collective effects in many body systems, despite recent advances~\cite{Seifert2012StochasticThermodynamics,Pietzonka2019AutonomousEngines}.
Understanding many body rectification can lead to methods to manipulate the flow of energy across materials without any imposed temperature biases~\cite{Zhu2016PersistentDirectional,Zhu2018TheoryManybody,Dubi2011ColloquiumHeat,Kanazawa2013HeatConduction,Martinez2017ColloidalHeat} and potentially facilitate the development of design principles for constructing synthetic molecular motor analogues. 
In this paper, we show how such rectification can be achieved in parity violating many body interacting systems. In particular, we show how a parity violating metamaterial~\cite{Nash2015TopologicalMechanics} can spontaneously rectify energy and motion in the absence of any imposed gradients, when it is allowed to interact with a bath of \textit{active} particles~\cite{Marchetti2013HydrodynamicsSoft,Koumakis2013TargetedDelivery,Woodhouse2018AutonomousActuation}. 

Our choice of parity violating metamaterial is inspired by a recent work where a metamaterial composed of interacting gyroscopes was introduced~\cite{Nash2015TopologicalMechanics}. 
These gyroscopic metamaterials were shown to support chiral topological edge modes whose origin lies in a violation of time reversal symmetry (TRS) in the microscopic equations of motion of the interacting gyroscopes~\cite{Nash2015TopologicalMechanics,Mitchell2018AmorphousTopological}. Crucially, the TRS violation is controlled by a combination of the spin of the gyroscopes and the geometry of the lattice. 

In this manuscript, we consider the effect of immersing such a TRS-violating metamaterial  in a bath of \textit{active} particles that violate the fluctuation dissipation relation~\cite{Fodor2016HowFar}. 
Our central result is a demonstration that this combination naturally rectifies stochastic fluctuations present in the bath. 
Crucially, the rectification mechanism relies on  both the TRS-violating properties of the metamaterial {\it and} the TRS violations implicit in the single particle fluctuations of the bath. 
Indeed, simply coupling the chiral topological metamaterial to a heat bath does not result in any fluxes or symmetry breaking on account of the Bohr-van Leeuwen theorem \cite{Pradhan2010NonexistenceClassical} that forbids any non-equilibrium currents in thermalized magnetic systems. 
Similarly, active baths coupled to passive metamaterials do not in general lead to rectification of fluctuations.

Our model system  generically supports a directed flux of energy across the metamaterial network. 
Unlike conventional energy flows, our energy flow does not require a temperature gradient. 
This flux can be routed through an otherwise isolated elastic object with the active network acting as a current source. 
Finally, we show that the microscopic mechanisms responsible for this energy flow can also potentially allow the {parity violating active} network to swim in and exert forces on a viscous fluid. 

We analytically and numerically demonstrate these results for a variety of network geometries. 
In particular, we construct an intuitive diagrammatic approach for computing the energy flux (and consequently the swim speed) that shows how our results can readily be applied to tailor flows in arbitrarily complex networks. 
Taken together, our results establish a new mechanism for rectification of energy, motion and forces in non-equilibrium parity violating metamaterials. Unlike  existing prescriptions, our mechanism exploits inherent asymmetries in the geometry and interactions of the material to achieve rectification.

The remainder of this paper is organized as follows:
In \secname~\ref{sec:model}, we introduce our model parity violating metamaterial and provide a microscopic definition for the energy flux.
In \secname~\ref{sec:linear_response}-\ref{sec:path} we analytically identify the ingredients for rectification of energy fluxes and construct a diagrammatic approach that reveals a relationship between energy flux and network geometries.
Finally in \secname~\ref{sec:passive}-\ref{sec:swimmer} we show that when the particles transmitting the energy flux are allowed to interact with a viscous low Reynolds number (Re) fluid, the nonreciprocal motions responsible for the energy flux can be utilized to pump the viscous fluid. 

\section{Model systems and energy flux} \label{sec:model}

\begin{figure}[tbp]
	\centering
	\includegraphics[width=0.48\textwidth]{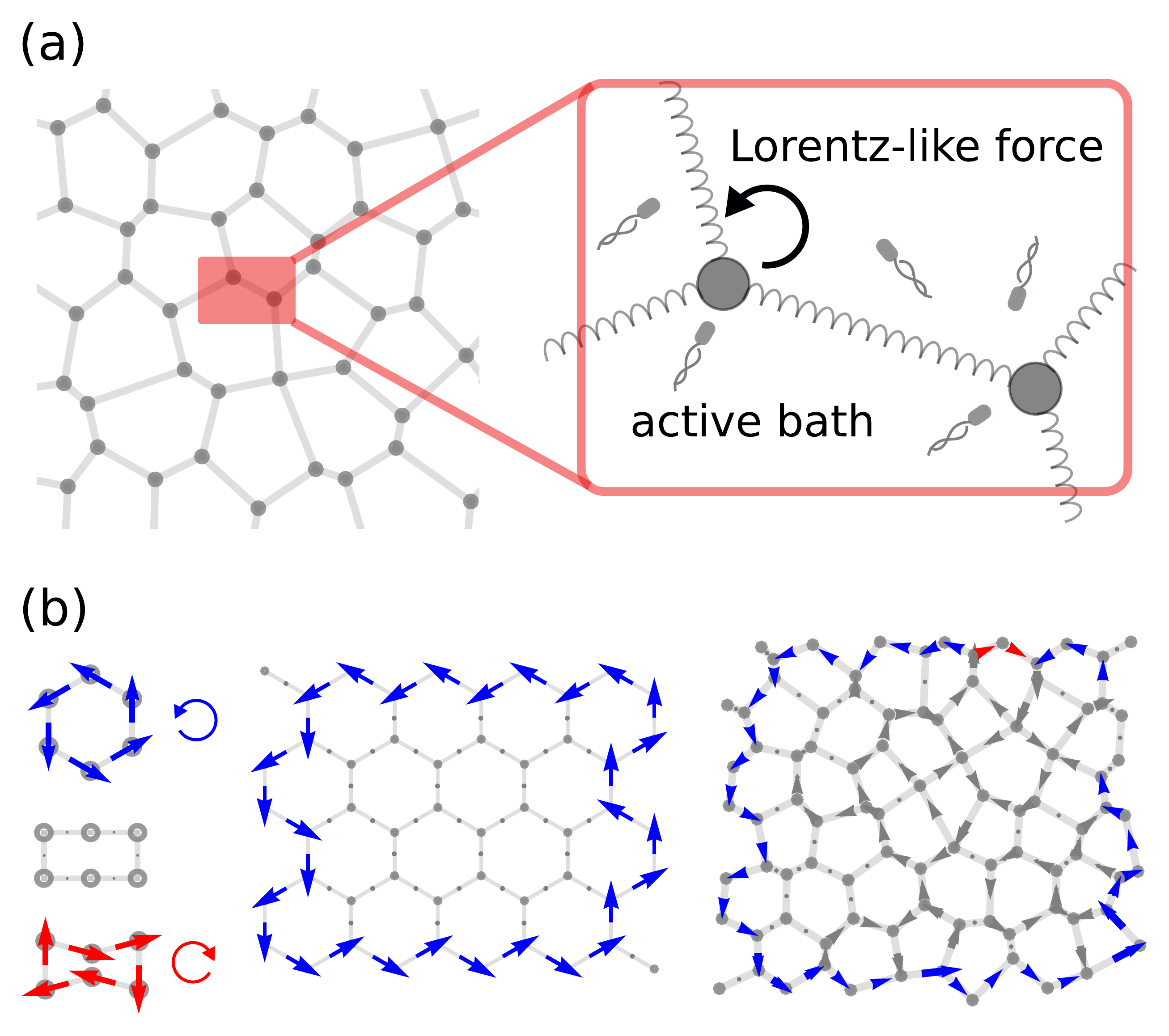}
    \caption{
    The model and the energy flux in exemplary networks.
    (a) Schematic of the model, a spring-mass network with Lorentz-like force and active bath on each particle.
    (b) Averaged energy flux from numerical calculations for a few typical networks. The flux direction and pattern can be controlled by the network geometry. In these figures, gray lines and dots represent the mechanical equilibrium structure of the network, and the size of the arrows is proportional to the magnitude of the averaged energy flux. The arrows are colored blue if it is counter-clockwise (CCW), red if clockwise (CW), and gray for fluxes not on the boundary. The numerical calculations were performed with all parameters set to $1$.
    }
    \label{fig:model_and_result}
\end{figure}

Our model TRS-violating metamaterial is a spring-mass network in which each particle feels a Lorentz-like force in addition to spring-like interactions with neighbors and  stochastic forcing from an active bath \cite{Fodor2016HowFar}  (\figurename~\ref{fig:model_and_result}a).
The equation of motion is:
\begin{equation} \label{eqn:GLE_single}
    m\dot{\bm{v}}_i = -k_g \bm{z}_i + \sum_j\bm{F}_{ji} + \bm{v}_i\times\bm{B} - \gamma\bm{v}_i + \bm{\eta}_i ,
\end{equation}
where $z_i \equiv \begin{pmatrix} x_i & y_i \end{pmatrix}^T$ is the displacement of particle $i$ from its mechanical equilibrium position, $F_{ji} = k (e_{ij}^T z_i + e_{ji}^T z_j) (-e_{ij})$, where $e_{ij}$ is the unit vector from the equilibrium position of $i$ to that of $j$, $\bm{B} = -B\mathbf{\hat{z}}$, and the  lorentz-force's customary electric charge-like factor is absorbed in $\bm{B}$.
The construction of our model system is motivated by the recently constructed topological gyroscopic metamaterials \cite{Nash2015TopologicalMechanics} and in the linearized regime, the equations of our model system are equivalent to the equations of motion of the gyroscopic metamaterials~\cite{Lee2018TopologicalDynamics}.

The last two terms describe the active bath, which consists of friction $-\gamma v_i$ and an Ornstein-Uhlenbeck (OU) colored noise $\eta_i$~\cite{Fodor2016HowFar}.
The colored noise has finite correlation time $\tau$ with statistics
\begin{equation} \label{eqn:noise_correlation}
    \expval{\eta_i(t)\eta_{j}^T(t')} = I\delta_{ij}\frac{\gamma T_a}{\tau} e^{-\frac{|t-t'|}{\tau}} ,
\end{equation}
where, for fixed $\tau$, the parameter $T_a$ controls the variance of the colored noise, and $I$ is the identity matrix with appropriate dimensions.
The time evolution of the OU noise can be described according to the following equation~\cite{Hanggi1994ColoredNoise},
\begin{equation} \label{eqn:noise_eom}
    \tau \dot{\eta}_i = -\eta_i + \sqrt{2\gamma T_a}\xi_i ,
\end{equation}
where $\xi_i$ is a delta function correlated white noise with unit variance.
The friction $-\gamma v_i$ and OU noise $\eta_i$ break fluctuation-dissipation relation, thus driving the system out of equilibrium~\cite{Fodor2016HowFar}.
The active bath reduces to a thermal equilibrium bath in the $\tau \rightarrow 0$ limit.

Since the particles in our model are tethered to their lattice sites, rectification of fluctuations, if any, does not result in any particle flows. Rather, rectified fluctuations can affect the transport of energy or heat. To study such phenomena, the observable we mainly focus on is the time-averaged energy flux between particles at steady state. For a system with pairwise interactions and on-site potentials, the energy flux $\expval{J_{ij}}$ from particle $i$ to $j$ reads
\begin{equation} \label{eqn:flux_def}
    \expval{J_{ij}} = \expval{\frac{1}{2} (\mathbf{v}_j \cdot \mathbf{F}_{ij} - \mathbf{v}_i \cdot \mathbf{F}_{ji})}
    = \expval{\mathbf{v}_j \cdot \mathbf{F}_{ij}}.
\end{equation}
To arrive at this formula, we first define the energy of a particle as the sum of its kinetic energy, on-site potential energy, and one half of the bond energies \cite{Lepri2003ThermalConduction}. Then we write down the energy balance relations using ideas from stochastic energetics~\cite{Sekimoto1998LangevinEquation}. Finally we identify the energy exchanged due to particle-particle interactions as the energy flux $\expval{J_{ij}}$ (derived in detail in the supplemental material \cite{SupplementalMaterial}).
We note that the energy flux can simply be interpreted as the rate at which work is done on particle $j$ by particle $i$.
Since this microscopic work is due to particles' stochastic motions, rather than due to an external control, the energy flux can also be interpreted as a heat flux \cite{Sekimoto1998LangevinEquation,Lepri2003ThermalConduction}. The averaged energy fluxes \eqnname~\eqref{eqn:flux_def} are identically equal to zero for a system at equilibrium with Boltzmann distribution.

Starting from the linear equations \eqnname~\eqref{eqn:GLE_single}, \eqref{eqn:noise_eom}, the energy fluxes can be solved numerically using methods introduced in \cite{Gardiner2009ItoCalculus,Ceriotti2010ColoredNoiseThermostats} (Supplemental Material \cite{SupplementalMaterial}).
A collection of numerical results are shown in \figurename~\ref{fig:model_and_result}b. We see nonzero energy rectification or energy fluxes can be generated in our chiral active system, and the flux direction or pattern changes with the network geometry.
Using a linear response theory, we now develop analytical expressions for the energy flux that 
reveal how a combination of chirality, non-equilibrium activity, and network geometry is responsible for generating energy fluxes.

\section{Linear response theory for energy flux} \label{sec:linear_response}
We begin by writing the equations of motion, \eqnname~\eqref{eqn:GLE_single}, in frequency space,
\begin{gather}
    \tilde{z}(\omega) = G^+(\omega) \tilde{\eta}(\omega), \label{eqn:response} \\
    G^{+}(\omega) \equiv [K + i\omega(\gamma I + BA) - m\omega^2I]^{-1}\,. \label{eqn:response_G}
\end{gather}
Here, we have represented the displacement of all particles using a column vector $z=\sum_i \ket{i}\otimes z_i$, with $\ket{i}$ denoting the 2D subspace of particle $i$. $\tilde{z}(\omega)$ and $\tilde{\eta}(\omega)$ denote the Fourier transform of $z$ and the OU noise $\eta$, respectively. $G^+$ is the response matrix, in which matrix $K$ encodes all on-site and spring forces $F$ according to $F=-Kz$, and $A$ is an antisymmetric matrix $A=\sum_i \ket{i}\bra{i}\otimes \mqty(0 & 1 \\ -1 & 0)$. \eqnname~\eqref{eqn:response} describes how the displacement responds to the noise.

Following the procedure in \cite{Kundu2011LargeDeviations}, the flux defined in \eqnname~\eqref{eqn:flux_def} can be expressed using $G^+$ as a spectral integral (Supplemental Material \cite{SupplementalMaterial})
\begin{align}
    \expval{J} &= \int_{-\infty}^\infty \dd{\omega} h(\omega) J^{FT}(\omega), \label{eqn:flux_integral} \\
    J^{FT}(\omega) &\equiv -\frac{T_a k}{2\pi} \Re \tr G^+(\omega)A^{as}, \\
    h(\omega) &= \frac{1}{1+\omega^2\tau^2},
\end{align}
where $A^{as}$ is an antisymmetric matrix
$A^{as} = -\ket{i}\bra{j} \otimes e_{ij}e_{ji}^T + \ket{j}\bra{i} \otimes e_{ji}e_{ij}^T$.
The response function $G^+(\omega)$ has no pole in the lower-half complex plane, but the colored noise introduces one pole at $\omega = -i/\tau$. Using the residue theorem we get a compact expression for the energy flux (Supplemental Material \cite{SupplementalMaterial})
\begin{equation} \label{eqn:flux_residue}
    \frac{\expval{J}}{T_a/\tau} = -\frac{k}{2} \tr G^+(-\frac{i}{\tau})A^{as} .
\end{equation}
\eqnname~\eqref{eqn:flux_integral} and \eqref{eqn:flux_residue} will serve as our starting point to understand the energy flux. While they contain all the information required to compute energy fluxes, they have limited utility as design principles. Indeed, as written down, they require that the flux be recomputed de novo for each new network geometry and non-equilibrium bath activity. In the next sections, we show that it is possible to expand \eqnname~\eqref{eqn:flux_integral} and \eqref{eqn:flux_residue} in forms that reveals design principles for controlling energy fluxes.

As an aside, one important property that can be directly obtained from a similar linear response analysis is that the energy fluxes satisfy Kirchoff's law, $\sum_i \expval{J_{ij}} = 0$. The Kirchoff's law shows that on average there is no energy exchange between particles and the active bath. 
To derive the Kirchoff's law, we calculate the average heat exchange between particle $i$ and the active bath $\expval{\bm{v}_i \cdot \bm{\eta}_i - \gamma \bm{v}_i \cdot \bm{v}_i}$, and following procedures in \cite{Kundu2011LargeDeviations}, this heat exchange can be shown to be zero (Supplemental Material \cite{SupplementalMaterial}).
The Kirchoff's law puts a strong constraint on possible energy flux patterns among particles, and some corollaries immediately follow, such as networks with no cycles cannot have nonzero flux, and fluxes of all bonds in a polygon network (as in \figurename~\ref{fig:model_and_result}b) are equal.

\section{Ingredients for energy rectification and their roles} \label{sec:fourier}

Compared with an ordinary thermal spring-mass network, that supports no energy fluxes in its equilibrium steady state, our parity violating metamaterial contains two extra components, the Lorentz force and the correlation in the noise.
We first show that these two components provide two necessary ingredients required to ensure energy rectification in our model. Next, we clarify the role played by the geometry of the network in controlling the energy flux.

\subsection{Lorentz force and non-equilibrium activity are necessary for the generation of an energy flux} 

\begin{figure}[tbp]
	\centering
	\includegraphics[width=0.48\textwidth]{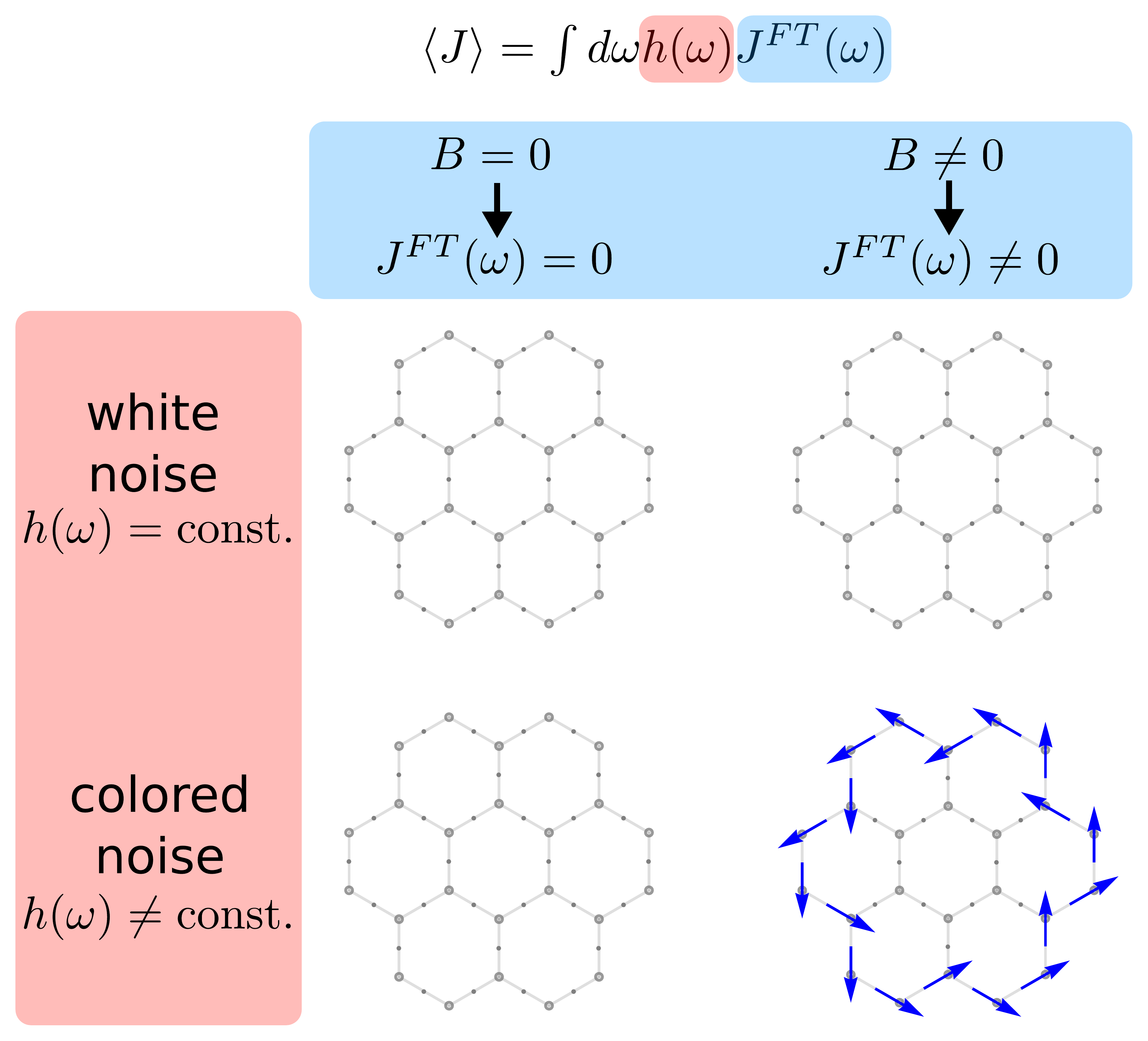}
    \caption{
    Necessary ingredients for generating nonzero energy fluxes.
    Both Lorentz-like force and colored noise are needed to generate nonzero fluxes.
    The role of Lorentz-like force is to provide chiral Fourier modes ($J^{FT}(\omega)\neq 0$). If $B=0$, $J^{FT}(\omega)$ is zero everywhere. 
    The role of colored noise is to provide weighted excitation $h(\omega)\neq \text{const.}$, which makes a non-vanishing averaged flux $\expval{J}$ possible.
    The numerical calculations were performed with $m,k_g,k,\gamma,T_a=1$. 
    }
    \label{fig:ingredients}
\end{figure}
We begin with \eqnname~\eqref{eqn:flux_integral} that represents the averaged flux in terms of functions $J^{FT}(\omega), h(\omega)$. In particular, we note that the function $J^{FT}(\omega)$ is proportional to the energy flux at Fourier frequency $\omega$ in an isolated damped variant of our network while the function $h(\omega)$ is proportional to the the noise spectrum, $\expval{\tilde{\eta}(\omega)\tilde{\eta}^*(\omega)} = 2\gamma T_a h(\omega) / t$.

To generate a nonzero flux, or equivalently make the integral nonzero in \eqnname~\eqref{eqn:flux_integral}, we need two requirements (\figurename~\ref{fig:ingredients}).
Firstly, $J^{FT}(\omega)$ should not be zero everywhere.
In the absence of a magnetic field, $B=0$, it can be easily shown from \eqnname~\eqref{eqn:flux_integral} that $J^{FT}(\omega)=0$. Indeed, the response function $G^+$ is symmetric or reciprocal in this case, and since $A^{as}$ is antisymmetric, the trace $\tr G^+(\omega) A^{as}=0$ at all values of $\omega$. Nonzero $B$ breaks the reciprocity of $G^+$, and can thus generate a nonzero $J^{FT}(\omega)$. 

Nonzero $J^{FT}(\omega)$ alone does not ensure a nonzero averaged energy flux, we further require that $h(\omega)$ not be constant.
Indeed, noise characterized by a constant $h(\omega)$ function corresponds to fluctuation dissipation preserving white noise. In such cases, our model system would be in equilibrium even in the presence of a magnetic field according to the Bohr-van Leeuwen theorem \cite{Pradhan2010NonexistenceClassical}.
A non-constant $h(\omega)$ function corresponding to a colored noise source, such as the fluctuation dissipation violating OU noise, $h(\omega)=1/(1+\omega^2\tau^2)$, can support a non-zero average energy flux. 

In summary, we see that $B$-field and a colored noise are two necessary ingredients to generate flux in our model chiral systems.
The role of the $B$-field is to break the reciprocity of response and generate Fourier modes such that $J^{FT}(\omega)\neq 0$. The role of the colored noise is to excite these modes in a weighted manner.

\subsection{Energy flux can be tuned as a function of lattice geometry} 
Apart from the these two ingredients, the geometry of the network also plays an important role.
Indeed in the small $\gamma$ regime, the existence of chiral modes with $J^{FT}(\omega)\neq 0$ can be heuristically explained by exploiting the connection between the slightly damped isolated variants of our system and the undamped isolated gyroscopic metamaterials~\cite{Nash2015TopologicalMechanics,Mitchell2018AmorphousTopological}.
Specifically, the slightly damped variant resonate near the eigen-frequencies of the undamped metamaterials, and hence exhibit Fourier modes that are close to the eigenmodes of the undamped system. Consequently, we infer that $J^{FT}(\omega)\neq 0$ as long as the corresponding eigenmodes in the undamped variant are chiral. As discussed in  \cite{Nash2015TopologicalMechanics}, the geometry of the network plays a crucial role in generating the chiral eigenmodes.
At larger $\gamma$'s, the Fourier modes of our damped isolated variant are no longer close to the chiral eigenmodes of gyroscopic metamaterials, but an emergent connection between eigenmodes and energy fluxes can be built (Supplemental Material \cite{SupplementalMaterial}).
In the next section, we further elaborate the role of geometry. Specifically, the central results of the next section provide compact expressions that elucidate the role played by geometry, Lorentz forces and non-equilibrium activity.

\section{Relationship between flux and network geometry} \label{sec:path}
\begin{figure}[tbp]
	\centering
	\includegraphics[width=0.48\textwidth]{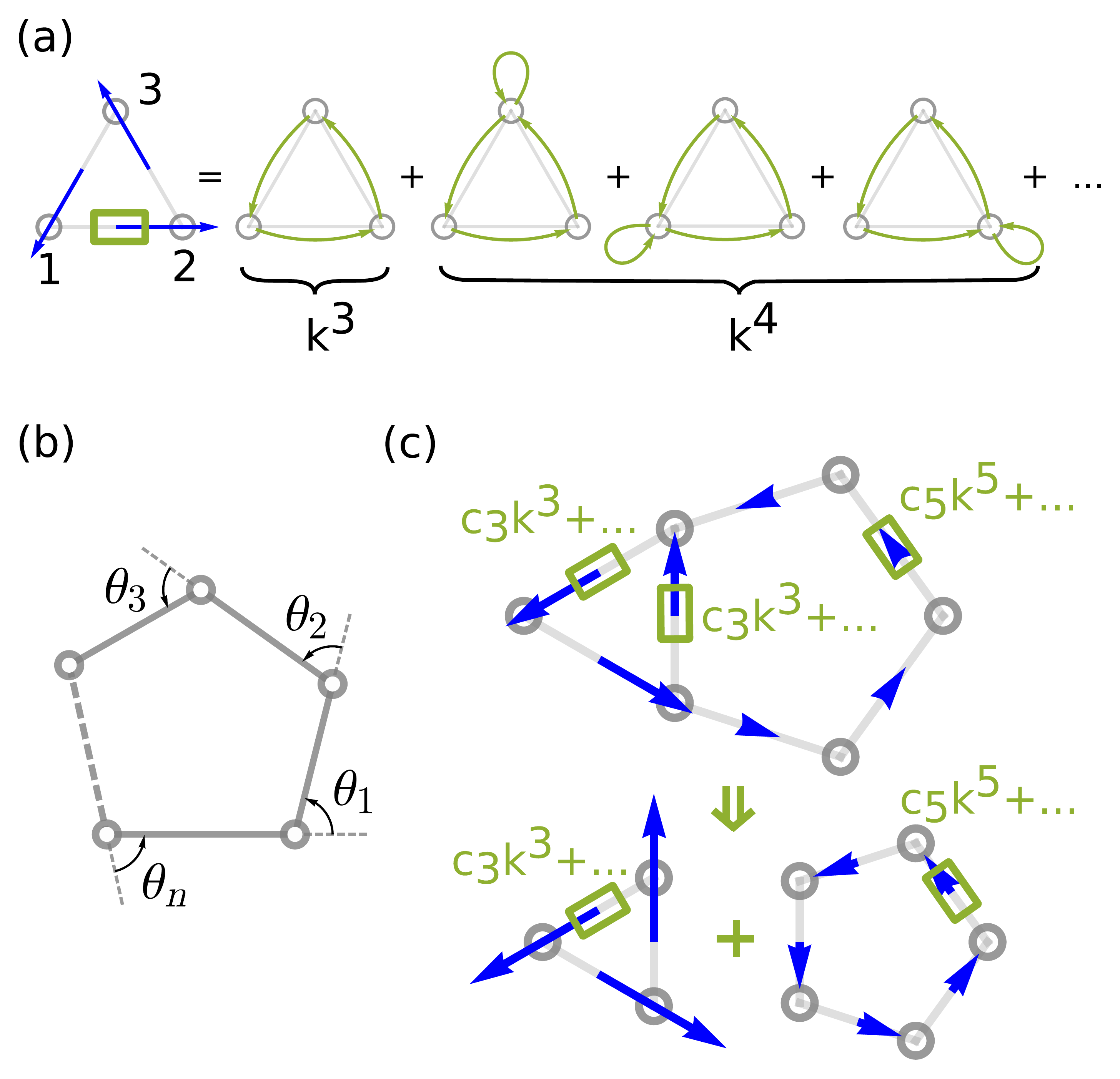}
    \caption{Illustrations of our diagrammatic technique.
    (a) Flux from $1$ to $2$ can be calculated by summing over paths. Each path is a pictorial representation of one term in the small-$k$ expansion, and is depicted using green arrows. The magnitude of path of length $n$ is on the order of $k^n$.
    (b) Schematic of a polygon path and its outer angles $\theta_1,\theta_2,\dots,\theta_n$. The flux of this path is simply \eqnname~\eqref{eqn:flux_path_polygon}.
    (c) For flux in complex networks, the leading order term is determined by the shortest cycles. Flux in the triangle part has order $k^3$, and the pre-factor $c_3$ is the same as that in a standalone triangle network. Likewise for the pentagon part. As a result, the flux in a complex network can be viewed as a combination of fluxes in its constituent cycles.
    }
    \label{fig:path_sum}
\end{figure}

One might expect, as is the case in ordinary Newtonian mechanical metamaterials, that the energy flux in our active system be simply given by considering the eigenmodes of the damped isolated system and adding the flux carried by each of these eigenmodes, weighted by their degree of excitation by the colored noise. This is however not the case. In our system, flux carried by a pair of excited eigenmodes is not equal to the sum of the fluxes carried by each eigenmode individually because the friction and the random forcing can induce cross-couplings between the different eigenvectors. While the flux can be described in terms of the eigenmodes of a modified isolated system, as described in the previous section, such a connection still does not provide an efficient framework for controlling and manipulating the energy fluxes using the network geometry and activity. 

In this section we develop a diagrammatic technique, which provides a simple intuitive method to compute energy fluxes and elucidates the relationship between flux and network geometry.
The diagrammatic technique is constructed by expanding the expressions for the energy flux \eqnname~\eqref{eqn:flux_residue} in small-$k$ regime and shows how the energy flux across a bond can be expressed as a sum over paths traversed along the network (\eqnname~\eqref{eqn:flux_path}). Our perturbation theory assigns a geometry dependent pre-factor for each path, thus elucidating the role played by network geometry in ensuring rectified energy fluxes (\eqnname~\eqref{eqn:flux_path_polygon}). Together, the central results of this section, summarized in  \eqnname~\eqref{eqn:flux_path}, \eqref{eqn:flux_path_polygon}, provide compact expressions that elucidate how geometry, $B$-field, and correlation time $\tau$ of the colored noise can combine to generate energy flows in networks with arbitrarily complex geometry and topologies. Crucially through this diagrammatic approach we demonstrate that the energy flux is controlled by the local geometry of the lattice. This result is surprising given the non-local dependence suggested by \eqnname~\eqref{eqn:flux_residue}. The diagrammatic techniques developed here enable the efficient control energy flux patterns in arbitrarily complex networks.

\subsection{Path summation and its rules}
Starting from the flux formula \eqnname~\eqref{eqn:flux_residue}, we expand the flux to different orders in the spring constant $k$. Then for each order, we further collect contributions from various paths.
In the final result, we write the total flux as a sum over the flux of paths (\figurename~\ref{fig:path_sum}a), (Supplemental Material \cite{SupplementalMaterial})
\begin{equation} \label{eqn:flux_path}
    \frac{\expval{J}}{T_a/\tau} = \sum_l J^\text{path}_l = \sum_l \frac{1}{2}(S_l - S_{-l}).
\end{equation}

The path rules are as follows.
For the flux from $i$ to $j$, valid paths are $l=i\rightarrow j\rightarrow l_3\rightarrow l_4\rightarrow \dots \rightarrow l_n\rightarrow i$, where $l_a$ and $l_b$ either has to be bonded or $l_a=l_b$. Paths that contain equal numbers of $i\rightarrow j$ and $j\rightarrow i$ do not contribute (e.g. path $i\rightarrow j\rightarrow i$), because either the path itself vanishes or it cancels with another path. As a result, paths appear as cycles.
The term $S_l$ is defined as $S_l \equiv (k/k_0)^n \tr R_\alpha (-K_s)_{i l_n} \cdots R_\alpha (-K_s)_{l_3j} R_\alpha (-K_s)_{ji}$.
In this definition, $k_0\equiv \sqrt{(k_g+\gamma/\tau+m/\tau^2)^2 + (B/\tau)^2}$ sets a characteristic scale for spring constant $k$. $R_\alpha \equiv \pmqty{\cos\alpha & -\sin\alpha \\ \sin\alpha & \cos\alpha}$ is a CCW rotation matrix. The angle $\alpha$ is defined as
\begin{equation} \label{eqn:path_alpha_def}
    \alpha \equiv \arcsin{\frac{B/\tau}{k_0}}.
\end{equation}
We note that all parameters ($m,k_g,k,B,\gamma,\tau$) are condensed into this single angle $\alpha$.
$(K_s)_{l_b l_a} \equiv \bra{l_b}(K-k_gI)\ket{l_a} / k$ is a $2\times 2$ submatrix used to calculate the non-dimensionalised spring force on $l_b$ due to the displacement of $l_a$.
$-l$ means $l$ in the reversed order.
The interval of convergence depends on the geometry of the whole network as well as the parameter $\alpha$. The typical value of the upper bound of $k/k_0$ ranges between $0.3$ and $0.6$.

The paths can be represented using diagrams, from which the flux $J^\text{path}_l$ can be calculated easily. For instance, the first diagram in \figurename~\ref{fig:path_sum}a represents the path $1\rightarrow 2\rightarrow 3\rightarrow 1$. To calculate $S_l$, one writes $(-K_s)_{l_bl_a}$ for each arrow $l_a\rightarrow l_b$, $R_\alpha$ for each node $l_a$, then multiply these matrices in the reversed order, and calculate the trace, e.g. $S_{1\rightarrow 2\rightarrow 3\rightarrow 1} = (k/k_0)^3 \tr R_\alpha (-K_s)_{13} R_\alpha (-K_s)_{32} R_\alpha (-K_s)_{21}$. To get $S_{-l}$, one takes the result of $S_l$ and replace $\alpha$ by $-\alpha$. Finally, $J^\text{path}_l$ can be calculated from the difference between $S_l$ and $S_{-l}$.

\subsection{Contributions to energy flux from leading-order paths}
Paths of length $n$ show up as terms of order $(k/k_0)^n$ in the energy flux expression. In the small $k$ regime, the main contribution to the flux comes from the lowest-order paths.
The usual lowest-order paths are polygonal cycles with no loops (loops are self-connecting edges like $l_a\rightarrow l_a$). For these polygonal paths, the flux formula \eqnname~\eqref{eqn:flux_path} reduces to a simple form (Supplemental Material \cite{SupplementalMaterial})
\begin{equation} \label{eqn:flux_path_polygon}
    J^\text{path}_\text{polygon} = \frac{1}{2} (\frac{k}{k_0})^n (\prod_i \cos(\theta_i - \alpha) - \prod_i \cos(\theta_i + \alpha)),
\end{equation}
where $\alpha$ is defined in \eqnname~\eqref{eqn:path_alpha_def}, $n$ is the number of nodes and $\theta_i$'s are outer angles (\figurename~\ref{fig:path_sum}b).
\eqnname~\eqref{eqn:flux_path_polygon} illustrates how geometry of the network, as characterized by the angles $\theta_i$, together with the condensed parameter $\alpha$ that encodes the nonreciprocity due to the $B$ field and the violation of fluctuation dissipation due to the colored noise, combine to generate energy fluxes.

For polygon networks, \eqnname~\eqref{eqn:flux_path_polygon} gives a direct relationship between the lowest-order flux and the network geometry. As an example, flux in an arbitrary triangle is $J \propto k^3 \sin\theta_1\sin\theta_2\sin\theta_3 + \mathcal{O}(k^4)$, whose $k^3$ term is always positive or CCW.
For complex networks, \eqnname~\eqref{eqn:flux_path_polygon} implys that its lowest-order flux can be viewed as a result of combining the flux of its constituent polygons, as illustrated in \figurename~\ref{fig:path_sum}c. This is because the flux of a polygonal path \eqnname~\eqref{eqn:flux_path_polygon} is not affected by any side chains on the nodes, and $J^\text{path}_\text{polygon}$ for a polygon in a complex network is the same as $J^\text{path}_\text{polygon}$ for the polygon when standalone.

Starting from the  diagrammatic expansion, the presence of localized energy fluxes in some networks (\figurename~\ref{fig:model_and_result}b) can be readily understood. Consider for instance the paths contributing to flux along an edge in a honeycomb-like network. Away from the boundary, the lower order path contributions to the energy flux cancel each other, and higher order path contributions become dominant. The size of the leading order path that contributes to the energy flux along a bond increases as a function of the distance of the bond from boundary. Such a scaling results in an exponential localization of the energy flux at the boundary of the network (Supplemental Material \cite{SupplementalMaterial}).

At the outset, given the long-ranged nature of elastic fluctuations, one might expect that the fluxes depend on topology in a highly non-local manner. Indeed, the expressions for flux in \eqnname~\eqref{eqn:flux_residue} suggest a complex non-local connection between the flux and the topology of the network. The results of this section show, however, that the flux can be in fact be controlled using effectively local rules. Hence, using the diagrammatic decomposition introduced here, it is easy to program energy flux patterns in networks of arbitrary complexity. Our diagrammatic expansion hence allows us to go beyond the need for de novo calculations suggested by our previous equations.

\section{Energy flux in a passive segment coupled to an active network} \label{sec:passive}
\label{sec:passiveflux}
\begin{figure}[tbp]
	\centering
	\includegraphics[width=0.48\textwidth]{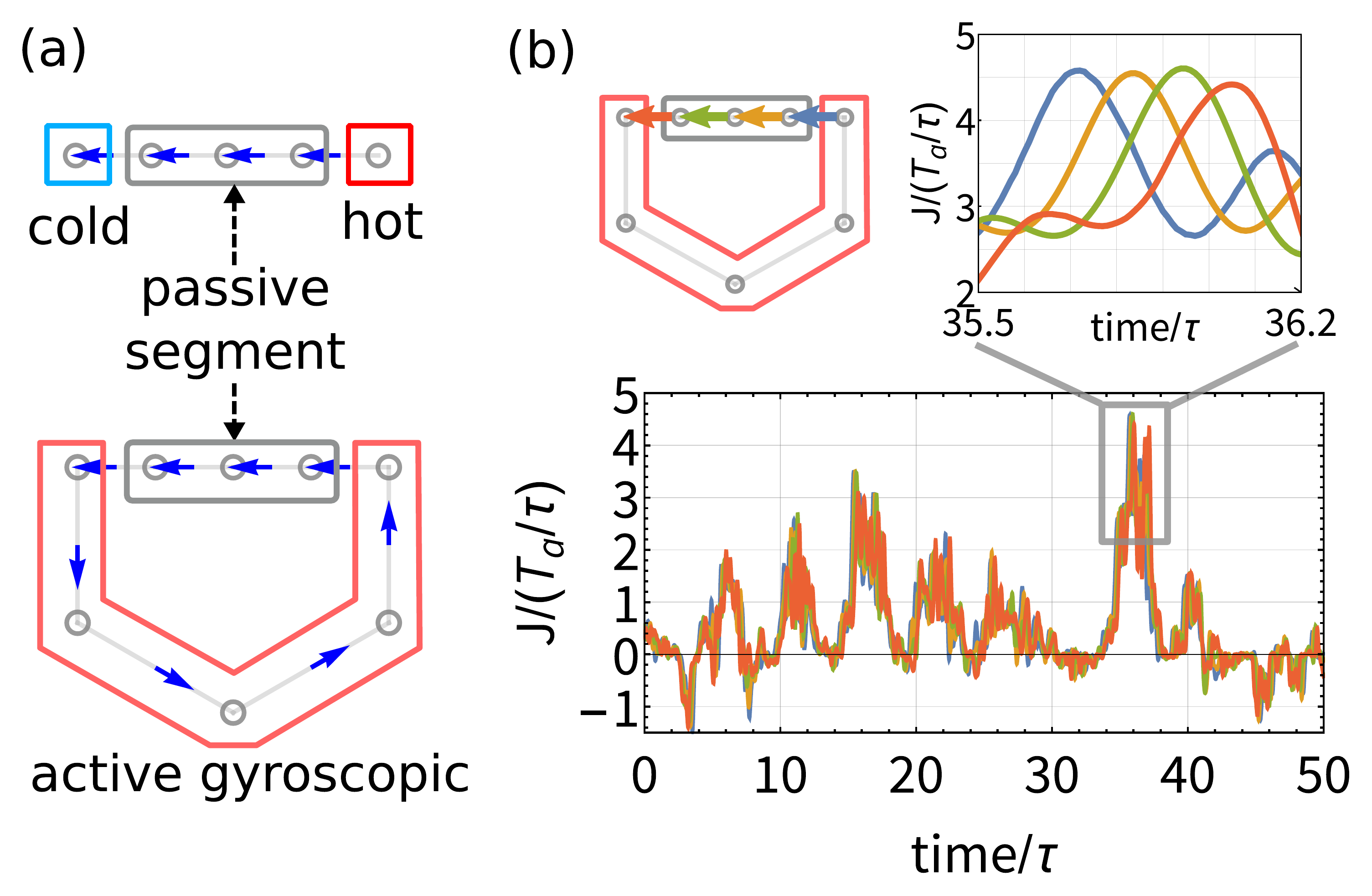}
    \caption{
    Driving energy through a passive chain.
    (a) Conventional energy transport in a passive material (boxed in gray) with temperature differences at two ends.
    Similarly, the active gyroscopic network can also drive energy flows through a passive segment.
    (b) Instantaneous energy flux $J$ through bonds in the passive segment from a simulation. Flux through different bonds are colored differently. $J$ is stochastic in general, however, during the period when $J$ is large, $J$ exhibits successive peaks in accordance with the direction of the flux.
    In the simulation setup, parameters for the active particles are: $m,k_g,\gamma=0.1, k=10, B,T_a,\tau=1$. Passive particles (boxed in gray) are constrained to 1D, and their $\gamma,T_a,k_g$ are set to $0$.
    }
    \label{fig:simulation}
\end{figure}

A canonical setup for the study of energy transport is  a passive material bar placed between tho heat reservoirs held at constant temperature (\figurename~\ref{fig:simulation}a). 
The generic result is that energy flows from the `hot' reservoir to the `cold' reservoir, and in the absence of a temperature difference there can be no net energy flux through the bar.

Placing a passive material bar \textendash in this case three masses not influenced by any magnetic fields and connected by springs in a linear geometry \textendash across a gap in our activated metamaterial, as illustrated in \figurename~\ref{fig:simulation}a, reveals a very different behavior: 
despite the absence of temperature gradients a persistent energy flux is measured through the passive material bar. 
From numerical calculations, the magnitude of the energy flux decays exponentially with the bar length. In the small $k$ limit, the flux in a bar with $n$ sites can be evaluated using the diagrammatic techniques developed in the previous section and is equal to $\expval{J}_n = \expval{J}_1 (k/(k_g+m/\tau^2))^{(n-1)}$. 
Using numerical simulations, we plot the instantaneous flux transmitted across the bonds on the passive segment in \figurename~\ref{fig:simulation}b (accompanying Supplemental Video \cite{SupplementalMaterial}). 
The instantaneous flux exhibits stochastic fluctuations. 
Large values of the instantaneous flux are transmitted across bonds sequentially in a wave like manner (\figurename~\ref{fig:simulation}b). This is reflected in the successive peaks in the instantaneous flux profile across the bonds of the passive segment.  The spacing between the peaks matches the sound speed in the passive chain. This result, in combination with the results of the previous sections, shows how one design active parity violating metamaterials that can act as energy pumps and support energy transport in passive materials even in the absence of any temperature gradients. 

Crucially, these results demonstrate how parity violating metamaterials can rectify non-equilibrium fluctuations.
In the following section, we consider whether these rectified fluctuations  when placed in contact with a viscous fluid, can act as low $Re$ swimmers or fluid pumps \cite{Taylor1951AnalysisSwimming,Purcell1977LifeLow,Golestanian2008AnalyticResults}. 

\section{Non-reciprocal motions responsible for energy fluxes can be used to generate forces} \label{sec:swimmer}

\begin{figure}[tbp]
	\centering
	\includegraphics[width=0.48\textwidth]{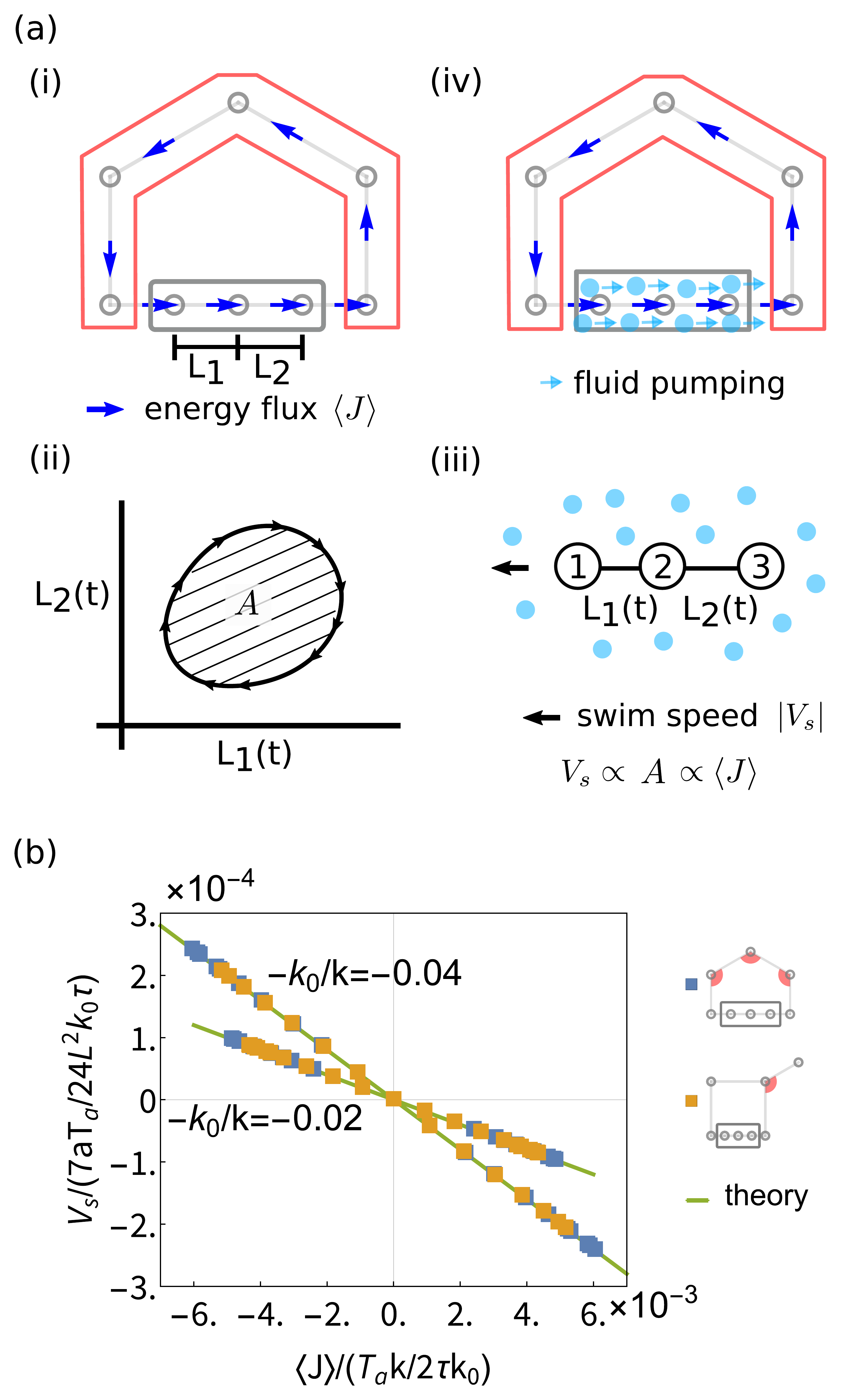}
    \caption{Utilization of non-equilibrium parity violating dynamics to power swimming and pumping in low Reynolds number ($Re$) media.
    (a) The energy flux in the active parity violating network is accompanied by nonreciprocal motions of the particles (i,ii). The nonreciprocal motion is a schematic for illustration purposes, and the real data is much more noisy. Using these nonreciprocal motions as input, a three bead linear object can be made to swim through a low Reynolds number ($Re$) medium (iii). Our analytical results in (b) below show that the swim speed $V_s$ is in fact proportional to the energy flux $\expval{J}$. (iv) Finally by immersing the passive segment into a low Reynolds fluid, the nonreciprocal motions can be used to pump the fluid. In this manner, the energy fluxes can be rectified for locomotion and force generation. 
    (b) Swim speed $V_s$ is proportional to the energy flux $\expval{J}$ in the active network. The proportionality constant is $-k_0/k$, which is independent of the network geometry. The series of dots for each $-k_0/k$ are obtained by varying the labelled angles (by red disk sectors) in pentagon networks or square+tail networks. The parameters chosen for the numerical calculations are $m,k_g=0.1$, $\tau=1$, $k=5$ for $k_0/k=0.04$ and $k=10$ for $k_0/k=0.02$. For the active part, $\gamma=0.1, B,T_a=1$. For the passive segment, $\gamma,B,T_a=0$.
    }
    \label{fig:swimmer}
\end{figure}

The rectification of energy has been our main focus so far. In this section, we show that it is possible to exploit the energy flux to rectify motions when our model systems are allowed to interact with a viscous fluid (\figurename~\ref{fig:swimmer}a).
We begin by considering the motion of the three masses in the passive material bar discussed in the previous section and in particular, consider the effect that their motion would have on a viscous fluid. We first do so by taking their recorded trajectories and asking whether three particles following these trajectories would `swim' in an external fluid.This calculation ignores any back action from the fluid on the dynamics of the segment. 
We then consider the effect of these forces in \secname~\ref{sec:pumping} and discuss regimes in which our energy conducting passive segment can generate forces when immersed in a viscous fluid. 
Together, these results demonstrate how a parity violating active metamaterial can be manipulated to exert forces and power motion in nanoscale materials.  

\subsection{Nonreciprocal motion as a swimming protocol}

In this section we consider a system of three spheres arranged in a linear configuration~\cite{Golestanian2008AnalyticResults} (\figurename~\ref{fig:swimmer}a(iii)), placed in a viscous fluid. 
This system is a minimal model for low Reynolds number swimming/pumping action. 
If the lengths of the two springs connecting the spheres, $L_1(t) = L+\Delta L_1(t), L_2(t) = L+\Delta L_2(t)$ are varied according to some prescribed protocol, the time-averaged swim speed is (\eqnname~(12) in \cite{Golestanian2008AnalyticResults})
\begin{equation} \label{eqn:swimmer_speed}
    V_s = \frac{7a}{24L^2} \expval{\Delta L_1 \dv{\Delta L_2}{t} - \dv{\Delta L_1}{t} \Delta{L_2}},
\end{equation}
where $a$ is the radius of the bead. Assumptions for this equation are $a/L \ll 1, \Delta L_i/L \ll 1$, and total external force on the swimmer is zero.

We now imagine recording the motions of the passive segment when it is connected to our parity violating metamaterial as in \secname~\ref{sec:passiveflux}(\figurename~\ref{fig:swimmer}a) and not coupled to a viscous fluid. 
This recorded motion can be used a protocol for modulating the configuration of an equivalent {\it swimmer} passive segment that is placed in a viscous fluid.
We compute the swim speed $V_s$ of the swimmer using \eqnname~\eqref{eqn:swimmer_speed} and find that it is in fact proportional to the energy flux, $\expval{J}$, conducted through the passive segment when it is coupled to the chiral active network (\figurename~\ref{fig:swimmer}b). The nonreciprocal motions that is responsible for energy fluxes can also be used as a protocol to generate motion in a low $Re$ fluid.
The proportionality constant between the swim speed, $V_s$ and the energy flux, $\expval{J}$, can be calculated using a modified diagrammatic technique (Supplemental Material \cite{SupplementalMaterial}),
\begin{equation} \label{eqn:swimmer_propto}
    \frac{V_s}{7a/24L^2} = -\frac{k_0}{k} \frac{\expval{J}}{k/2},
\end{equation}
where $k_0 = k_g + m/\tau^2$ ($B,\gamma=0$ for the passive segment).
This result \eqnname~\eqref{eqn:swimmer_propto} holds beyond small-$k$ regime because all orders of paths are considered.
\figurename~\ref{fig:swimmer}b and \eqnname~\eqref{eqn:swimmer_propto} together establish that one can relate the swim speed to the flux of energy in the parity violating metamaterial. 
Similar proportionality between $V_s$ and $\expval{J}$ can be expected for other types of three-sphere swimmers, such as one where one sphere is much larger than the other two \cite{Golestanian2008ThreesphereLowReynoldsnumber}. This is because the swim speed is generically proportional to the area enclosed in the $\Delta L_i$ space \cite{Golestanian2009StochasticLow}. This area is also proportional to the energy flux $\expval{J}$ (Supplemental Material \cite{SupplementalMaterial}).

\subsection{Force generation in a viscous medium using the rectified energy fluxes}
\label{sec:pumping}
Finally, we now consider the scenario where the passive segment, which is connected to the active network, is immersed in a fluid (\figurename~\ref{fig:swimmer}a(iv)). Since the passive segment is tethered due to $k_g$ and its connection to the active network, it cannot swim indefinitely. However, a stalled swimmer can potentially pump the fluid \cite{Leoni2009BasicSwimmer}. We now analyse the regime for parameters where this pumping is possible.
In order to utilize the connection between the swim speed and energy flux, we require that the fluid minimally perturbs the dynamics of the passive segment. Such a regime requires the dissipation rate due to the viscous fluid to be much smaller than the energy flux through the segment. This condition can be expressed as $\eta_f a v^2 \ll J$, where $v$ is the characteristic velocity of a bead in the passive segment, $\eta_f$ is the dynamic viscosity of the fluid, and $J$ is the energy flux in the absence of the fluid.
In addition, the constraint of low Reynolds number requires that $Re = \rho_f a v /\eta_f \ll 1$, where $\rho_f$ is fluid's density.
Writing these two conditions together, the requirement reads $J \gg \eta_f a v^2 \gg \rho_f a^2 v^3$. 
Here, we have ignored hydrodynamic interactions between the beads, because it is a higher-order perturbation with the order of $a/L$. 

As a numerical example, this condition can be satisfied by setting $k=5\times 10^{-5} kg/s^2, k_g=1\times 10^{-6} kg/s^2$ for springs (values of optical traps \cite{Leoni2009BasicSwimmer}), $a=10^{-6}m$ for all beads (size used in \cite{Leoni2009BasicSwimmer}), $T_a=10^{-18}J, \tau=1s$ for the active bath \cite{Wu2000ParticleDiffusion}, $\rho_f=10^3kg/m^3, \eta_f=10^{-3}kg/(m\cdot s)$ for liquid (water), and $B=10^{-5} kg/s$ for the $B$-field.
For these parameters, we obtain $v=3.8\times 10^{-6}m/s$, and the three relevant scales $J=5\times 10^{-19}J/s, \eta_f av^2=1.4\times 10^{-22}J/s, a^2v^3\rho_f=5.4\times 10^{-29}J/s$, which indeed satisfy $J \gg \eta_f a v^2 \gg \rho_f a^2 v^3$.
If the separation between beads is $L=10^{-7}m$, the swim speed calculated from \eqnname~\eqref{eqn:swimmer_propto} is $1\times 10^{-8} m/s$. 
The swim speed can provide an estimate of the potential speed at which the fluid can be pumped due to the energy fluxes. 
This speed can be scaled up by increasing $T_a$.
We note the extremely high value of the $B$ field required to generate a sizable averaged flux $\langle J \rangle$. To make practical use of this model, we anticipate that it will become necessary to instead consider models with interacting gyroscopes~\cite{Nash2015TopologicalMechanics} or Coriolis forces \cite{Kahlert2012MagnetizingComplex}. Large $B$-field or Lorentz force analogues are easier to achieve in these cases. 

In summary, the results of this section show that the energy fluxes generated by our active parity violating metamaterial can be used to rectify motion and generate forces on the nanoscale. Our calculations also show how the forces generated by our metamaterial are in fact proportional to the energy fluxes. Hence, together with the results from the previous sections that show how energy fluxes in arbitrarily complex active metamaterial networks can be controlled, our results provide a broad framework to generate and modulate forces using active parity violating metamaterials.

\section{Conclusion} \label{sec:conclusion}

In conclusion, we have established a general set of design principles for rectifying energy and motion in non-equilibrium parity violating metamaterials. In particular, our central results show how a combination of time reversal symmetry violation due to the interactions and Lorentz forces in the metamaterial, and due to the non-equilibrium fluctuations of the active bath, can result in a general strategy for rectification of energy and motion. Extending these ideas to non-equilibrium parity violating metamaterials with non-linear interactions or materials composed of active chiral particles can potentially lead to new strategies for the construction of synthetic molecular motor analogues. These ideas will be considered in future work.

\section*{Acknowledgements}
This work was primarily supported by NSF DMR- MRSEC 1420709. SV acknowledges support from the Sloan Foundation, startup funds from the University of Chicago and support from the National Science Foundation under award number DMR-1848306. WTMI Acknowledges support from NSF EFRI NewLAW grant 1741685 and the Packard foundation. 


\begin{thebibliography}{36}%
\makeatletter
\providecommand \@ifxundefined [1]{%
    \@ifx{#1\undefined}
}%
\providecommand \@ifnum [1]{%
    \ifnum #1\expandafter \@firstoftwo
    \else \expandafter \@secondoftwo
    \fi
}%
\providecommand \@ifx [1]{%
    \ifx #1\expandafter \@firstoftwo
    \else \expandafter \@secondoftwo
    \fi
}%
\providecommand \natexlab [1]{#1}%
\providecommand \enquote  [1]{``#1''}%
\providecommand \bibnamefont  [1]{#1}%
\providecommand \bibfnamefont [1]{#1}%
\providecommand \citenamefont [1]{#1}%
\providecommand \href@noop [0]{\@secondoftwo}%
\providecommand \href [0]{\begingroup \@sanitize@url \@href}%
\providecommand \@href[1]{\@@startlink{#1}\@@href}%
\providecommand \@@href[1]{\endgroup#1\@@endlink}%
\providecommand \@sanitize@url [0]{\catcode `\\12\catcode `\$12\catcode
    `\&12\catcode `\#12\catcode `\^12\catcode `\_12\catcode `\%12\relax}%
\providecommand \@@startlink[1]{}%
\providecommand \@@endlink[0]{}%
\providecommand \url  [0]{\begingroup\@sanitize@url \@url }%
\providecommand \@url [1]{\endgroup\@href {#1}{\urlprefix }}%
\providecommand \urlprefix  [0]{URL }%
\providecommand \Eprint [0]{\href }%
\providecommand \doibase [0]{http://dx.doi.org/}%
\providecommand \selectlanguage [0]{\@gobble}%
\providecommand \bibinfo  [0]{\@secondoftwo}%
\providecommand \bibfield  [0]{\@secondoftwo}%
\providecommand \translation [1]{[#1]}%
\providecommand \BibitemOpen [0]{}%
\providecommand \bibitemStop [0]{}%
\providecommand \bibitemNoStop [0]{.\EOS\space}%
\providecommand \EOS [0]{\spacefactor3000\relax}%
\providecommand \BibitemShut  [1]{\csname bibitem#1\endcsname}%
\let\auto@bib@innerbib\@empty
\bibitem [{\citenamefont
    {Seifert}(2012)}]{Seifert2012StochasticThermodynamics}%
    \BibitemOpen
    \bibfield  {author} {\bibinfo {author} {\bibfnamefont {Udo}\ \bibnamefont
    {Seifert}},\ }\bibfield  {title} {\enquote {\bibinfo {title} {Stochastic
    thermodynamics, fluctuation theorems and molecular machines.}}\ }\href
    {\doibase 10.1088/0034-4885/75/12/126001} {\bibfield  {journal} {\bibinfo
    {journal} {Reports on progress in physics. Physical Society (Great Britain)}\
    }\textbf {\bibinfo {volume} {75}},\ \bibinfo {pages} {126001} (\bibinfo
    {year} {2012})},\ \Eprint {http://arxiv.org/abs/1205.4176v1}
    {arXiv:1205.4176v1} \BibitemShut {NoStop}%
\bibitem [{\citenamefont {Coskun}\ \emph {et~al.}(2011)\citenamefont {Coskun},
    \citenamefont {Banaszak}, \citenamefont {Astumian}, \citenamefont
    {Stoddart},\ and\ \citenamefont {Grzybowski}}]{Coskun2011GreatExpectations}%
    \BibitemOpen
    \bibfield  {author} {\bibinfo {author} {\bibfnamefont {Ali}\ \bibnamefont
    {Coskun}}, \bibinfo {author} {\bibfnamefont {Michal}\ \bibnamefont
    {Banaszak}}, \bibinfo {author} {\bibfnamefont {R.~Dean}\ \bibnamefont
    {Astumian}}, \bibinfo {author} {\bibfnamefont {J.~Fraser}\ \bibnamefont
    {Stoddart}}, \ and\ \bibinfo {author} {\bibfnamefont {Bartosz~A.}\
    \bibnamefont {Grzybowski}},\ }\bibfield  {title} {\enquote {\bibinfo {title}
    {Great expectations: Can artificial molecular machines deliver on their
    promise?}}\ }\href {\doibase 10.1039/C1CS15262A} {\bibfield  {journal}
    {\bibinfo  {journal} {Chemical Society Reviews}\ }\textbf {\bibinfo {volume}
    {41}},\ \bibinfo {pages} {19--30} (\bibinfo {year} {2011})}\BibitemShut
    {NoStop}%
\bibitem [{\citenamefont {Jarzynski}\ and\ \citenamefont
    {Mazonka}(1999)}]{Jarzynski1999FeynmanRatchet}%
    \BibitemOpen
    \bibfield  {author} {\bibinfo {author} {\bibfnamefont {C}~\bibnamefont
    {Jarzynski}}\ and\ \bibinfo {author} {\bibfnamefont {O}~\bibnamefont
    {Mazonka}},\ }\bibfield  {title} {\enquote {\bibinfo {title} {Feynman's
    ratchet and pawl: An exactly solvable model.}}\ }\href {\doibase
    10.1103/PhysRevE.59.6448} {\bibfield  {journal} {\bibinfo  {journal}
    {Physical review. E, Statistical physics, plasmas, fluids, and related
    interdisciplinary topics}\ }\textbf {\bibinfo {volume} {59}},\ \bibinfo
    {pages} {6448--6459} (\bibinfo {year} {1999})}\BibitemShut {NoStop}%
\bibitem [{\citenamefont {Mogilner}\ and\ \citenamefont
    {Oster}(1996)}]{Mogilner1996CellMotility}%
    \BibitemOpen
    \bibfield  {author} {\bibinfo {author} {\bibfnamefont {A.}~\bibnamefont
    {Mogilner}}\ and\ \bibinfo {author} {\bibfnamefont {G.}~\bibnamefont
    {Oster}},\ }\bibfield  {title} {\enquote {\bibinfo {title} {Cell motility
    driven by actin polymerization},}\ }\href {\doibase
    10.1016/S0006-3495(96)79496-1} {\bibfield  {journal} {\bibinfo  {journal}
    {Biophysical Journal}\ }\textbf {\bibinfo {volume} {71}},\ \bibinfo {pages}
    {3030--3045} (\bibinfo {year} {1996})}\BibitemShut {NoStop}%
\bibitem [{\citenamefont {Chernyak}\ and\ \citenamefont
    {Sinitsyn}(2008)}]{Chernyak2008PumpingRestriction}%
    \BibitemOpen
    \bibfield  {author} {\bibinfo {author} {\bibfnamefont {V.~Y.}\ \bibnamefont
    {Chernyak}}\ and\ \bibinfo {author} {\bibfnamefont {N.~A.}\ \bibnamefont
    {Sinitsyn}},\ }\bibfield  {title} {\enquote {\bibinfo {title} {Pumping
    {{Restriction Theorem}} for {{Stochastic Networks}}},}\ }\href {\doibase
    10.1103/PhysRevLett.101.160601} {\bibfield  {journal} {\bibinfo  {journal}
    {Physical Review Letters}\ }\textbf {\bibinfo {volume} {101}},\ \bibinfo
    {pages} {160601} (\bibinfo {year} {2008})}\BibitemShut {NoStop}%
\bibitem [{\citenamefont {Rahav}\ \emph {et~al.}(2008)\citenamefont {Rahav},
    \citenamefont {Horowitz},\ and\ \citenamefont
    {Jarzynski}}]{Rahav2008DirectedFlow}%
    \BibitemOpen
    \bibfield  {author} {\bibinfo {author} {\bibfnamefont {Saar}\ \bibnamefont
    {Rahav}}, \bibinfo {author} {\bibfnamefont {Jordan}\ \bibnamefont
    {Horowitz}}, \ and\ \bibinfo {author} {\bibfnamefont {Christopher}\
    \bibnamefont {Jarzynski}},\ }\bibfield  {title} {\enquote {\bibinfo {title}
    {Directed flow in nonadiabatic stochastic pumps},}\ }\href {\doibase
    10.1103/PhysRevLett.101.140602} {\bibfield  {journal} {\bibinfo  {journal}
    {Physical Review Letters}\ }\textbf {\bibinfo {volume} {101}},\ \bibinfo
    {pages} {1--4} (\bibinfo {year} {2008})},\ \Eprint
    {http://arxiv.org/abs/0808.0015} {arXiv:0808.0015} \BibitemShut {NoStop}%
\bibitem [{\citenamefont {Sinitsyn}\ and\ \citenamefont
    {Nemenman}(2007)}]{Sinitsyn2007UniversalGeometric}%
    \BibitemOpen
    \bibfield  {author} {\bibinfo {author} {\bibfnamefont {N.~A.}\ \bibnamefont
    {Sinitsyn}}\ and\ \bibinfo {author} {\bibfnamefont {Ilya}\ \bibnamefont
    {Nemenman}},\ }\bibfield  {title} {\enquote {\bibinfo {title} {Universal
    geometric theory of mesoscopic stochastic pumps and reversible ratchets},}\
    }\href {\doibase 10.1103/PhysRevLett.99.220408} {\bibfield  {journal}
    {\bibinfo  {journal} {Physical Review Letters}\ }\textbf {\bibinfo {volume}
    {99}},\ \bibinfo {pages} {1--4} (\bibinfo {year} {2007})},\ \Eprint
    {http://arxiv.org/abs/0705.2057} {arXiv:0705.2057} \BibitemShut {NoStop}%
\bibitem [{\citenamefont {Pietzonka}\ \emph {et~al.}(2019)\citenamefont
    {Pietzonka}, \citenamefont {Fodor}, \citenamefont {Lohrmann}, \citenamefont
    {Cates},\ and\ \citenamefont {Seifert}}]{Pietzonka2019AutonomousEngines}%
    \BibitemOpen
    \bibfield  {author} {\bibinfo {author} {\bibfnamefont {Patrick}\ \bibnamefont
    {Pietzonka}}, \bibinfo {author} {\bibfnamefont {{\'E}tienne}\ \bibnamefont
    {Fodor}}, \bibinfo {author} {\bibfnamefont {Christoph}\ \bibnamefont
    {Lohrmann}}, \bibinfo {author} {\bibfnamefont {Michael~E.}\ \bibnamefont
    {Cates}}, \ and\ \bibinfo {author} {\bibfnamefont {Udo}\ \bibnamefont
    {Seifert}},\ }\bibfield  {title} {\enquote {\bibinfo {title} {Autonomous
    engines driven by active matter: {{Energetics}} and design principles},}\
    }\href@noop {} {\bibfield  {journal} {\bibinfo  {journal} {arXiv:1905.00373
    [cond-mat]}\ } (\bibinfo {year} {2019})},\ \Eprint
    {http://arxiv.org/abs/1905.00373} {arXiv:1905.00373 [cond-mat]} \BibitemShut
    {NoStop}%
\bibitem [{\citenamefont {Zhu}\ and\ \citenamefont
    {Fan}(2016)}]{Zhu2016PersistentDirectional}%
    \BibitemOpen
    \bibfield  {author} {\bibinfo {author} {\bibfnamefont {Linxiao}\ \bibnamefont
    {Zhu}}\ and\ \bibinfo {author} {\bibfnamefont {Shanhui}\ \bibnamefont
    {Fan}},\ }\bibfield  {title} {\enquote {\bibinfo {title} {Persistent
    {{Directional Current}} at {{Equilibrium}} in {{Nonreciprocal Many}}-{{Body
    Near Field Electromagnetic Heat Transfer}}},}\ }\href {\doibase
    10.1103/PhysRevLett.117.134303} {\bibfield  {journal} {\bibinfo  {journal}
    {Physical Review Letters}\ }\textbf {\bibinfo {volume} {117}},\ \bibinfo
    {pages} {1--6} (\bibinfo {year} {2016})},\ \Eprint
    {http://arxiv.org/abs/1602.08454} {arXiv:1602.08454} \BibitemShut {NoStop}%
\bibitem [{\citenamefont {Zhu}\ \emph {et~al.}(2018)\citenamefont {Zhu},
    \citenamefont {Guo},\ and\ \citenamefont {Fan}}]{Zhu2018TheoryManybody}%
    \BibitemOpen
    \bibfield  {author} {\bibinfo {author} {\bibfnamefont {Linxiao}\ \bibnamefont
    {Zhu}}, \bibinfo {author} {\bibfnamefont {Yu}~\bibnamefont {Guo}}, \ and\
    \bibinfo {author} {\bibfnamefont {Shanhui}\ \bibnamefont {Fan}},\ }\bibfield
    {title} {\enquote {\bibinfo {title} {Theory of many-body radiative heat
    transfer without the constraint of reciprocity},}\ }\href {\doibase
    10.1103/PhysRevB.97.094302} {\bibfield  {journal} {\bibinfo  {journal}
    {Physical Review B}\ }\textbf {\bibinfo {volume} {97}},\ \bibinfo {pages}
    {1--11} (\bibinfo {year} {2018})}\BibitemShut {NoStop}%
\bibitem [{\citenamefont {Dubi}\ and\ \citenamefont
    {Di~Ventra}(2011)}]{Dubi2011ColloquiumHeat}%
    \BibitemOpen
    \bibfield  {author} {\bibinfo {author} {\bibfnamefont {Yonatan}\ \bibnamefont
    {Dubi}}\ and\ \bibinfo {author} {\bibfnamefont {Massimiliano}\ \bibnamefont
    {Di~Ventra}},\ }\bibfield  {title} {\enquote {\bibinfo {title} {Colloquium:
    {{Heat}} flow and thermoelectricity in atomic and molecular junctions},}\
    }\href {\doibase 10.1103/RevModPhys.83.131} {\bibfield  {journal} {\bibinfo
    {journal} {Reviews of Modern Physics}\ }\textbf {\bibinfo {volume} {83}},\
    \bibinfo {pages} {131--155} (\bibinfo {year} {2011})}\BibitemShut {NoStop}%
\bibitem [{\citenamefont {Kanazawa}\ \emph {et~al.}(2013)\citenamefont
    {Kanazawa}, \citenamefont {Sagawa},\ and\ \citenamefont
    {Hayakawa}}]{Kanazawa2013HeatConduction}%
    \BibitemOpen
    \bibfield  {author} {\bibinfo {author} {\bibfnamefont {Kiyoshi}\ \bibnamefont
    {Kanazawa}}, \bibinfo {author} {\bibfnamefont {Takahiro}\ \bibnamefont
    {Sagawa}}, \ and\ \bibinfo {author} {\bibfnamefont {Hisao}\ \bibnamefont
    {Hayakawa}},\ }\bibfield  {title} {\enquote {\bibinfo {title} {Heat
    conduction induced by non-{{Gaussian}} athermal fluctuations},}\ }\href
    {\doibase 10.1103/PhysRevE.87.052124} {\bibfield  {journal} {\bibinfo
    {journal} {Physical Review E}\ }\textbf {\bibinfo {volume} {87}},\ \bibinfo
    {pages} {052124} (\bibinfo {year} {2013})},\ \Eprint
    {http://arxiv.org/abs/1209.2222v3} {arXiv:1209.2222v3} \BibitemShut {NoStop}%
\bibitem [{\citenamefont {Mart\'inez}\ \emph {et~al.}(2017)\citenamefont
    {Mart\'inez}, \citenamefont {Rold{\'a}n}, \citenamefont {Dinis},\ and\
    \citenamefont {Rica}}]{Martinez2017ColloidalHeat}%
    \BibitemOpen
    \bibfield  {author} {\bibinfo {author} {\bibfnamefont {Ignacio~A.}\
    \bibnamefont {Mart\'inez}}, \bibinfo {author} {\bibfnamefont {{\'E}dgar}\
    \bibnamefont {Rold{\'a}n}}, \bibinfo {author} {\bibfnamefont {Luis}\
    \bibnamefont {Dinis}}, \ and\ \bibinfo {author} {\bibfnamefont
    {Ra{\'u}l~Alberto}\ \bibnamefont {Rica}},\ }\bibfield  {title} {\enquote
    {\bibinfo {title} {Colloidal heat engines: A review},}\ }\href {\doibase
    10.1039/C6SM00923A} {\bibfield  {journal} {\bibinfo  {journal} {Soft Matter}\
    }\textbf {\bibinfo {volume} {13}},\ \bibinfo {pages} {22--36} (\bibinfo
    {year} {2017})}\BibitemShut {NoStop}%
\bibitem [{\citenamefont {Nash}\ \emph {et~al.}(2015)\citenamefont {Nash},
    \citenamefont {Kleckner}, \citenamefont {Read}, \citenamefont {Vitelli},
    \citenamefont {Turner},\ and\ \citenamefont
    {Irvine}}]{Nash2015TopologicalMechanics}%
    \BibitemOpen
    \bibfield  {author} {\bibinfo {author} {\bibfnamefont {Lisa~M.}\ \bibnamefont
    {Nash}}, \bibinfo {author} {\bibfnamefont {Dustin}\ \bibnamefont {Kleckner}},
    \bibinfo {author} {\bibfnamefont {Alismari}\ \bibnamefont {Read}}, \bibinfo
    {author} {\bibfnamefont {Vincenzo}\ \bibnamefont {Vitelli}}, \bibinfo
    {author} {\bibfnamefont {Ari~M.}\ \bibnamefont {Turner}}, \ and\ \bibinfo
    {author} {\bibfnamefont {William T.~M.}\ \bibnamefont {Irvine}},\ }\bibfield
    {title} {\enquote {\bibinfo {title} {Topological mechanics of gyroscopic
    metamaterials},}\ }\href {\doibase 10.1073/pnas.1507413112} {\bibfield
    {journal} {\bibinfo  {journal} {Proceedings of the National Academy of
    Sciences}\ }\textbf {\bibinfo {volume} {112}},\ \bibinfo {pages}
    {14495--14500} (\bibinfo {year} {2015})},\ \Eprint
    {http://arxiv.org/abs/1504.03362} {arXiv:1504.03362} \BibitemShut {NoStop}%
\bibitem [{\citenamefont {Marchetti}\ \emph {et~al.}(2013)\citenamefont
    {Marchetti}, \citenamefont {Joanny}, \citenamefont {Ramaswamy}, \citenamefont
    {Liverpool}, \citenamefont {Prost}, \citenamefont {Rao},\ and\ \citenamefont
    {Simha}}]{Marchetti2013HydrodynamicsSoft}%
    \BibitemOpen
    \bibfield  {author} {\bibinfo {author} {\bibfnamefont {M.~C.}\ \bibnamefont
    {Marchetti}}, \bibinfo {author} {\bibfnamefont {J.~F.}\ \bibnamefont
    {Joanny}}, \bibinfo {author} {\bibfnamefont {S.}~\bibnamefont {Ramaswamy}},
    \bibinfo {author} {\bibfnamefont {T.~B.}\ \bibnamefont {Liverpool}}, \bibinfo
    {author} {\bibfnamefont {J.}~\bibnamefont {Prost}}, \bibinfo {author}
    {\bibfnamefont {Madan}\ \bibnamefont {Rao}}, \ and\ \bibinfo {author}
    {\bibfnamefont {R.~Aditi}\ \bibnamefont {Simha}},\ }\bibfield  {title}
    {\enquote {\bibinfo {title} {Hydrodynamics of soft active matter},}\ }\href
    {\doibase 10.1103/RevModPhys.85.1143} {\bibfield  {journal} {\bibinfo
    {journal} {Reviews of Modern Physics}\ }\textbf {\bibinfo {volume} {85}},\
    \bibinfo {pages} {1143--1189} (\bibinfo {year} {2013})},\ \Eprint
    {http://arxiv.org/abs/1207.2929} {arXiv:1207.2929} \BibitemShut {NoStop}%
\bibitem [{\citenamefont {Koumakis}\ \emph {et~al.}(2013)\citenamefont
    {Koumakis}, \citenamefont {Lepore}, \citenamefont {Maggi},\ and\
    \citenamefont {Di~Leonardo}}]{Koumakis2013TargetedDelivery}%
    \BibitemOpen
    \bibfield  {author} {\bibinfo {author} {\bibfnamefont {N.}~\bibnamefont
    {Koumakis}}, \bibinfo {author} {\bibfnamefont {A.}~\bibnamefont {Lepore}},
    \bibinfo {author} {\bibfnamefont {C.}~\bibnamefont {Maggi}}, \ and\ \bibinfo
    {author} {\bibfnamefont {R.}~\bibnamefont {Di~Leonardo}},\ }\bibfield
    {title} {\enquote {\bibinfo {title} {Targeted delivery of colloids by
    swimming bacteria},}\ }\href {\doibase 10.1038/ncomms3588} {\bibfield
    {journal} {\bibinfo  {journal} {Nature Communications}\ }\textbf {\bibinfo
    {volume} {4}},\ \bibinfo {pages} {1--6} (\bibinfo {year} {2013})}\BibitemShut
    {NoStop}%
\bibitem [{\citenamefont {Woodhouse}\ \emph {et~al.}(2018)\citenamefont
    {Woodhouse}, \citenamefont {Ronellenfitsch},\ and\ \citenamefont
    {Dunkel}}]{Woodhouse2018AutonomousActuation}%
    \BibitemOpen
    \bibfield  {author} {\bibinfo {author} {\bibfnamefont {Francis~G.}\
    \bibnamefont {Woodhouse}}, \bibinfo {author} {\bibfnamefont {Henrik}\
    \bibnamefont {Ronellenfitsch}}, \ and\ \bibinfo {author} {\bibfnamefont
    {J{\"o}rn}\ \bibnamefont {Dunkel}},\ }\bibfield  {title} {\enquote {\bibinfo
    {title} {Autonomous {{Actuation}} of {{Zero Modes}} in {{Mechanical Networks
    Far}} from {{Equilibrium}}},}\ }\href {\doibase
    10.1103/PhysRevLett.121.178001} {\bibfield  {journal} {\bibinfo  {journal}
    {Physical Review Letters}\ }\textbf {\bibinfo {volume} {121}},\ \bibinfo
    {pages} {178001} (\bibinfo {year} {2018})},\ \Eprint
    {http://arxiv.org/abs/1805.07728} {arXiv:1805.07728} \BibitemShut {NoStop}%
\bibitem [{\citenamefont {Mitchell}\ \emph {et~al.}(2018)\citenamefont
    {Mitchell}, \citenamefont {Nash}, \citenamefont {Hexner}, \citenamefont
    {Turner},\ and\ \citenamefont {Irvine}}]{Mitchell2018AmorphousTopological}%
    \BibitemOpen
    \bibfield  {author} {\bibinfo {author} {\bibfnamefont {Noah~P.}\ \bibnamefont
    {Mitchell}}, \bibinfo {author} {\bibfnamefont {Lisa~M.}\ \bibnamefont
    {Nash}}, \bibinfo {author} {\bibfnamefont {Daniel}\ \bibnamefont {Hexner}},
    \bibinfo {author} {\bibfnamefont {Ari~M.}\ \bibnamefont {Turner}}, \ and\
    \bibinfo {author} {\bibfnamefont {William T.~M.}\ \bibnamefont {Irvine}},\
    }\bibfield  {title} {\enquote {\bibinfo {title} {Amorphous topological
    insulators constructed from random point sets},}\ }\href {\doibase
    10.1038/s41567-017-0024-5} {\bibfield  {journal} {\bibinfo  {journal} {Nature
    Physics}\ } (\bibinfo {year} {2018}),\ 10.1038/s41567-017-0024-5}\BibitemShut
    {NoStop}%
\bibitem [{\citenamefont {Fodor}\ \emph {et~al.}(2016)\citenamefont {Fodor},
    \citenamefont {Nardini}, \citenamefont {Cates}, \citenamefont {Tailleur},
    \citenamefont {Visco},\ and\ \citenamefont {{van
    Wijland}}}]{Fodor2016HowFar}%
    \BibitemOpen
    \bibfield  {author} {\bibinfo {author} {\bibfnamefont {{\'E}tienne}\
    \bibnamefont {Fodor}}, \bibinfo {author} {\bibfnamefont {Cesare}\
    \bibnamefont {Nardini}}, \bibinfo {author} {\bibfnamefont {Michael~E.}\
    \bibnamefont {Cates}}, \bibinfo {author} {\bibfnamefont {Julien}\
    \bibnamefont {Tailleur}}, \bibinfo {author} {\bibfnamefont {Paolo}\
    \bibnamefont {Visco}}, \ and\ \bibinfo {author} {\bibfnamefont
    {Fr{\'e}d{\'e}ric}\ \bibnamefont {{van Wijland}}},\ }\bibfield  {title}
    {\enquote {\bibinfo {title} {How {{Far}} from {{Equilibrium Is Active
    Matter}}?}}\ }\href {\doibase 10.1103/PhysRevLett.117.038103} {\bibfield
    {journal} {\bibinfo  {journal} {Physical Review Letters}\ }\textbf {\bibinfo
    {volume} {117}},\ \bibinfo {pages} {038103} (\bibinfo {year} {2016})},\
    \Eprint {http://arxiv.org/abs/1604.00953} {arXiv:1604.00953} \BibitemShut
    {NoStop}%
\bibitem [{\citenamefont {Pradhan}\ and\ \citenamefont
    {Seifert}(2010)}]{Pradhan2010NonexistenceClassical}%
    \BibitemOpen
    \bibfield  {author} {\bibinfo {author} {\bibfnamefont {P.}~\bibnamefont
    {Pradhan}}\ and\ \bibinfo {author} {\bibfnamefont {U.}~\bibnamefont
    {Seifert}},\ }\bibfield  {title} {\enquote {\bibinfo {title} {Nonexistence of
    classical diamagnetism and nonequilibrium fluctuation theorems for charged
    particles on a curved surface},}\ }\href {\doibase
    10.1209/0295-5075/89/37001} {\bibfield  {journal} {\bibinfo  {journal} {EPL
    (Europhysics Letters)}\ }\textbf {\bibinfo {volume} {89}},\ \bibinfo {pages}
    {37001} (\bibinfo {year} {2010})},\ \Eprint {http://arxiv.org/abs/0912.4697}
    {arXiv:0912.4697} \BibitemShut {NoStop}%
\bibitem [{\citenamefont {Lee}\ \emph {et~al.}(2018)\citenamefont {Lee},
    \citenamefont {Li}, \citenamefont {Jin}, \citenamefont {Liu},\ and\
    \citenamefont {Zhang}}]{Lee2018TopologicalDynamics}%
    \BibitemOpen
    \bibfield  {author} {\bibinfo {author} {\bibfnamefont {Ching~Hua}\
    \bibnamefont {Lee}}, \bibinfo {author} {\bibfnamefont {Guangjie}\
    \bibnamefont {Li}}, \bibinfo {author} {\bibfnamefont {Guliuxin}\ \bibnamefont
    {Jin}}, \bibinfo {author} {\bibfnamefont {Yuhan}\ \bibnamefont {Liu}}, \ and\
    \bibinfo {author} {\bibfnamefont {Xiao}\ \bibnamefont {Zhang}},\ }\bibfield
    {title} {\enquote {\bibinfo {title} {Topological dynamics of gyroscopic and
    {{Floquet}} lattices from {{Newton}}'s laws},}\ }\href {\doibase
    10.1103/PhysRevB.97.085110} {\bibfield  {journal} {\bibinfo  {journal}
    {Physical Review B}\ }\textbf {\bibinfo {volume} {97}},\ \bibinfo {pages}
    {085110} (\bibinfo {year} {2018})},\ \Eprint
    {http://arxiv.org/abs/1701.03385} {arXiv:1701.03385} \BibitemShut {NoStop}%
\bibitem [{\citenamefont {Hanggi}\ and\ \citenamefont
    {Jung}(1994)}]{Hanggi1994ColoredNoise}%
    \BibitemOpen
    \bibfield  {author} {\bibinfo {author} {\bibfnamefont {Peter}\ \bibnamefont
    {Hanggi}}\ and\ \bibinfo {author} {\bibfnamefont {Peter}\ \bibnamefont
    {Jung}},\ }\bibfield  {title} {\enquote {\bibinfo {title} {Colored {{Noise}}
    in {{Dynamical Systems}}},}\ }in\ \href {\doibase 10.1002/9780470141489.ch4}
    {\emph {\bibinfo {booktitle} {Advances in {{Chemical Physics}}}}}\ (\bibinfo
    {publisher} {{John Wiley \& Sons, Ltd}},\ \bibinfo {year} {1994})\ pp.\
    \bibinfo {pages} {239--326}\BibitemShut {NoStop}%
\bibitem [{\citenamefont {Lepri}(2003)}]{Lepri2003ThermalConduction}%
    \BibitemOpen
    \bibfield  {author} {\bibinfo {author} {\bibfnamefont {Stefano}\ \bibnamefont
    {Lepri}},\ }\bibfield  {title} {\enquote {\bibinfo {title} {Thermal
    conduction in classical low-dimensional lattices},}\ }\href {\doibase
    10.1016/S0370-1573(02)00558-6} {\bibfield  {journal} {\bibinfo  {journal}
    {Physics Reports}\ }\textbf {\bibinfo {volume} {377}},\ \bibinfo {pages}
    {1--80} (\bibinfo {year} {2003})}\BibitemShut {NoStop}%
\bibitem [{\citenamefont {Sekimoto}(1998)}]{Sekimoto1998LangevinEquation}%
    \BibitemOpen
    \bibfield  {author} {\bibinfo {author} {\bibfnamefont {Ken}\ \bibnamefont
    {Sekimoto}},\ }\bibfield  {title} {\enquote {\bibinfo {title} {Langevin
    {{Equation}} and {{Thermodynamics}}},}\ }\href {\doibase 10.1143/PTPS.130.17}
    {\bibfield  {journal} {\bibinfo  {journal} {Progress of Theoretical Physics
    Supplement}\ }\textbf {\bibinfo {volume} {130}},\ \bibinfo {pages} {17--27}
    (\bibinfo {year} {1998})}\BibitemShut {NoStop}%
\bibitem [{Sup()}]{SupplementalMaterial}%
    \BibitemOpen
    \href@noop {} {}\bibinfo {note} {See Supplemental Material at [URL will be
    inserted by publisher]}\BibitemShut {NoStop}%
\bibitem [{\citenamefont {Gardiner}(2009)}]{Gardiner2009ItoCalculus}%
    \BibitemOpen
    \bibfield  {author} {\bibinfo {author} {\bibfnamefont {Crispin}\ \bibnamefont
    {Gardiner}},\ }\bibfield  {title} {\enquote {\bibinfo {title} {The {{Ito
    Calculus}} and {{Stochastic Differential Equations}}},}\ }in\ \href@noop {}
    {\emph {\bibinfo {booktitle} {Stochastic {{Methods}}}}}\ (\bibinfo
    {publisher} {{Springer-Verlag Berlin Heidelberg}},\ \bibinfo {year} {2009})\
    \bibinfo {edition} {4th}\ ed.,\ Chap.~\bibinfo {chapter} {4}, p.\ \bibinfo
    {pages} {107}\BibitemShut {NoStop}%
\bibitem [{\citenamefont {Ceriotti}\ \emph {et~al.}(2010)\citenamefont
    {Ceriotti}, \citenamefont {Bussi},\ and\ \citenamefont
    {Parrinello}}]{Ceriotti2010ColoredNoiseThermostats}%
    \BibitemOpen
    \bibfield  {author} {\bibinfo {author} {\bibfnamefont {Michele}\ \bibnamefont
    {Ceriotti}}, \bibinfo {author} {\bibfnamefont {Giovanni}\ \bibnamefont
    {Bussi}}, \ and\ \bibinfo {author} {\bibfnamefont {Michele}\ \bibnamefont
    {Parrinello}},\ }\bibfield  {title} {\enquote {\bibinfo {title}
    {Colored-{{Noise Thermostats}} {\`a} la {{Carte}}},}\ }\href {\doibase
    10.1021/ct900563s} {\bibfield  {journal} {\bibinfo  {journal} {Journal of
    Chemical Theory and Computation}\ }\textbf {\bibinfo {volume} {6}},\ \bibinfo
    {pages} {1170--1180} (\bibinfo {year} {2010})}\BibitemShut {NoStop}%
\bibitem [{\citenamefont {Kundu}\ \emph {et~al.}(2011)\citenamefont {Kundu},
    \citenamefont {Sabhapandit},\ and\ \citenamefont
    {Dhar}}]{Kundu2011LargeDeviations}%
    \BibitemOpen
    \bibfield  {author} {\bibinfo {author} {\bibfnamefont {Anupam}\ \bibnamefont
    {Kundu}}, \bibinfo {author} {\bibfnamefont {Sanjib}\ \bibnamefont
    {Sabhapandit}}, \ and\ \bibinfo {author} {\bibfnamefont {Abhishek}\
    \bibnamefont {Dhar}},\ }\bibfield  {title} {\enquote {\bibinfo {title} {Large
    deviations of heat flow in harmonic chains},}\ }\href {\doibase
    10.1088/1742-5468/2011/03/P03007} {\bibfield  {journal} {\bibinfo  {journal}
    {Journal of Statistical Mechanics: Theory and Experiment}\ }\textbf {\bibinfo
    {volume} {2011}} (\bibinfo {year} {2011}),\
    10.1088/1742-5468/2011/03/P03007},\ \Eprint {http://arxiv.org/abs/1101.3669}
    {arXiv:1101.3669} \BibitemShut {NoStop}%
\bibitem [{\citenamefont {Taylor}(1951)}]{Taylor1951AnalysisSwimming}%
    \BibitemOpen
    \bibfield  {author} {\bibinfo {author} {\bibfnamefont {G.}~\bibnamefont
    {Taylor}},\ }\bibfield  {title} {\enquote {\bibinfo {title} {Analysis of the
    {{Swimming}} of {{Microscopic Organisms}}},}\ }\href {\doibase
    10.1098/rspa.1951.0218} {\bibfield  {journal} {\bibinfo  {journal}
    {Proceedings of the Royal Society A: Mathematical, Physical and Engineering
    Sciences}\ }\textbf {\bibinfo {volume} {209}},\ \bibinfo {pages} {447--461}
    (\bibinfo {year} {1951})},\ \Eprint {http://arxiv.org/abs/0912.1431}
    {arXiv:0912.1431} \BibitemShut {NoStop}%
\bibitem [{\citenamefont {Purcell}(1977)}]{Purcell1977LifeLow}%
    \BibitemOpen
    \bibfield  {author} {\bibinfo {author} {\bibfnamefont {E.~M.}\ \bibnamefont
    {Purcell}},\ }\bibfield  {title} {\enquote {\bibinfo {title} {Life at low
    {{Reynolds}} number},}\ }\href {\doibase 10.1119/1.10903} {\bibfield
    {journal} {\bibinfo  {journal} {American Journal of Physics}\ }\textbf
    {\bibinfo {volume} {45}},\ \bibinfo {pages} {3--11} (\bibinfo {year}
    {1977})},\ \Eprint {http://arxiv.org/abs/1011.1669v3} {arXiv:1011.1669v3}
    \BibitemShut {NoStop}%
\bibitem [{\citenamefont {Golestanian}\ and\ \citenamefont
    {Ajdari}(2008)}]{Golestanian2008AnalyticResults}%
    \BibitemOpen
    \bibfield  {author} {\bibinfo {author} {\bibfnamefont {Ramin}\ \bibnamefont
    {Golestanian}}\ and\ \bibinfo {author} {\bibfnamefont {Armand}\ \bibnamefont
    {Ajdari}},\ }\bibfield  {title} {\enquote {\bibinfo {title} {Analytic results
    for the three-sphere swimmer at low {{Reynolds}} number},}\ }\href {\doibase
    10.1103/PhysRevE.77.036308} {\bibfield  {journal} {\bibinfo  {journal}
    {Physical Review E - Statistical, Nonlinear, and Soft Matter Physics}\
    }\textbf {\bibinfo {volume} {77}},\ \bibinfo {pages} {1--6} (\bibinfo {year}
    {2008})},\ \Eprint {http://arxiv.org/abs/0711.3700} {arXiv:0711.3700}
    \BibitemShut {NoStop}%
\bibitem [{\citenamefont
    {Golestanian}(2008)}]{Golestanian2008ThreesphereLowReynoldsnumber}%
    \BibitemOpen
    \bibfield  {author} {\bibinfo {author} {\bibfnamefont {R.}~\bibnamefont
    {Golestanian}},\ }\bibfield  {title} {\enquote {\bibinfo {title}
    {Three-sphere low-{{Reynolds}}-number swimmer with a cargo container},}\
    }\href {\doibase 10.1140/epje/i2007-10276-2} {\bibfield  {journal} {\bibinfo
    {journal} {The European Physical Journal E}\ }\textbf {\bibinfo {volume}
    {25}},\ \bibinfo {pages} {1--4} (\bibinfo {year} {2008})}\BibitemShut
    {NoStop}%
\bibitem [{\citenamefont {Golestanian}\ and\ \citenamefont
    {Ajdari}(2009)}]{Golestanian2009StochasticLow}%
    \BibitemOpen
    \bibfield  {author} {\bibinfo {author} {\bibfnamefont {Ramin}\ \bibnamefont
    {Golestanian}}\ and\ \bibinfo {author} {\bibfnamefont {Armand}\ \bibnamefont
    {Ajdari}},\ }\bibfield  {title} {\enquote {\bibinfo {title} {Stochastic low
    {{Reynolds}} number swimmers},}\ }\href {\doibase
    10.1088/0953-8984/21/20/204104} {\bibfield  {journal} {\bibinfo  {journal}
    {Journal of Physics Condensed Matter}\ }\textbf {\bibinfo {volume} {21}}
    (\bibinfo {year} {2009}),\ 10.1088/0953-8984/21/20/204104},\ \Eprint
    {http://arxiv.org/abs/0901.1624} {arXiv:0901.1624} \BibitemShut {NoStop}%
\bibitem [{\citenamefont {Leoni}\ \emph {et~al.}(2009)\citenamefont {Leoni},
    \citenamefont {Kotar}, \citenamefont {Bassetti}, \citenamefont {Cicuta},\
    and\ \citenamefont {Lagomarsino}}]{Leoni2009BasicSwimmer}%
    \BibitemOpen
    \bibfield  {author} {\bibinfo {author} {\bibfnamefont {Marco}\ \bibnamefont
    {Leoni}}, \bibinfo {author} {\bibfnamefont {Jurij}\ \bibnamefont {Kotar}},
    \bibinfo {author} {\bibfnamefont {Bruno}\ \bibnamefont {Bassetti}}, \bibinfo
    {author} {\bibfnamefont {Pietro}\ \bibnamefont {Cicuta}}, \ and\ \bibinfo
    {author} {\bibfnamefont {Marco~Cosentino}\ \bibnamefont {Lagomarsino}},\
    }\bibfield  {title} {\enquote {\bibinfo {title} {A basic swimmer at low
    {{Reynolds}} number},}\ }\href {\doibase 10.1039/B812393D} {\bibfield
    {journal} {\bibinfo  {journal} {Soft Matter}\ }\textbf {\bibinfo {volume}
    {5}},\ \bibinfo {pages} {472--476} (\bibinfo {year} {2009})},\ \Eprint
    {http://arxiv.org/abs/0807.1867} {arXiv:0807.1867} \BibitemShut {NoStop}%
\bibitem [{\citenamefont {Wu}\ and\ \citenamefont
    {Libchaber}(2000)}]{Wu2000ParticleDiffusion}%
    \BibitemOpen
    \bibfield  {author} {\bibinfo {author} {\bibfnamefont {Xiao-Lun}\
    \bibnamefont {Wu}}\ and\ \bibinfo {author} {\bibfnamefont {Albert}\
    \bibnamefont {Libchaber}},\ }\bibfield  {title} {\enquote {\bibinfo {title}
    {Particle {{Diffusion}} in a {{Quasi}}-{{Two}}-{{Dimensional Bacterial
    Bath}}},}\ }\href {\doibase 10.1103/PhysRevLett.84.3017} {\bibfield
    {journal} {\bibinfo  {journal} {Physical Review Letters}\ }\textbf {\bibinfo
    {volume} {84}},\ \bibinfo {pages} {3017--3020} (\bibinfo {year}
    {2000})}\BibitemShut {NoStop}%
\bibitem [{\citenamefont {K{\"a}hlert}\ \emph {et~al.}(2012)\citenamefont
    {K{\"a}hlert}, \citenamefont {Carstensen}, \citenamefont {Bonitz},
    \citenamefont {L{\"o}wen}, \citenamefont {Greiner},\ and\ \citenamefont
    {Piel}}]{Kahlert2012MagnetizingComplex}%
    \BibitemOpen
    \bibfield  {author} {\bibinfo {author} {\bibfnamefont {H.}~\bibnamefont
    {K{\"a}hlert}}, \bibinfo {author} {\bibfnamefont {J.}~\bibnamefont
    {Carstensen}}, \bibinfo {author} {\bibfnamefont {M.}~\bibnamefont {Bonitz}},
    \bibinfo {author} {\bibfnamefont {H.}~\bibnamefont {L{\"o}wen}}, \bibinfo
    {author} {\bibfnamefont {F.}~\bibnamefont {Greiner}}, \ and\ \bibinfo
    {author} {\bibfnamefont {A.}~\bibnamefont {Piel}},\ }\bibfield  {title}
    {\enquote {\bibinfo {title} {Magnetizing a {{Complex Plasma}} without a
    {{Magnetic Field}}},}\ }\href {\doibase 10.1103/PhysRevLett.109.155003}
    {\bibfield  {journal} {\bibinfo  {journal} {Physical Review Letters}\
    }\textbf {\bibinfo {volume} {109}},\ \bibinfo {pages} {155003} (\bibinfo
    {year} {2012})}\BibitemShut {NoStop}%
\end{thebibliography}
%

\end{document}


\newcommand{\eqnname}{Eq.}
\newcommand{\secname}{Sec.}
\renewcommand{\theequation}{S\arabic{equation}}
\renewcommand{\thefigure}{S\arabic{figure}}
\renewcommand{\bibnumfmt}[1]{[S#1]}
\renewcommand{\citenumfont}[1]{S#1}

\title{Supplemental Material: Rectification of energy and motion in non-equilibrium parity violating metamaterials}

\author{Zhenghan Liao}
\affiliation{Department of Chemistry, University of Chicago, Chicago, IL, 60637, USA}
\author{William T. M. Irvine}
\affiliation{Department of Physics, University of Chicago, Chicago, IL 60637, USA}
\affiliation{James Franck Institute, University of Chicago, Chicago, IL 60637, USA}
\affiliation{ Enrico Fermi Institute, University of Chicago, Chicago, IL, 60637, USA}
\author{Suriyanarayanan Vaikuntanathan}
\affiliation{Department of Chemistry, University of Chicago, Chicago, IL, 60637, USA}
\affiliation{James Franck Institute, University of Chicago, Chicago, IL 60637, USA}

\maketitle


\section{The formula for energy flux}
In this section, we derive the formula for energy flux, \eqnname~(4) in the main text.
The force $F$ in this section is a general conservative force, which does not need to be linear.
We use the following strategy to determine the energy flux. First we define the energy $E_i$ of particle $i$, and then write down an energy balance relation, that expresses the infinitesimal energy change $\dd E_i$ using stochastic calculus. Finally, we collect terms in $\dd E_i$ that couple neighboring particles together and identify this as the energy transfer among particles.

The energy of particle $i$ is defined as
\begin{equation} \label{eqnS:energy_individual}
E_i = \frac{1}{2}m_iv_i^Tv_i + U_{ii} + \frac{1}{2}\sum_{j\neq i}U_{ij},
\end{equation}
where the first term is the kinetic energy, the second term denotes the on-site potential, and the last term is the shared spring energy between the particle and its neighbors.

We use Ito's formula to calculate $\dd E_i$. 
 Ito calculus provides the advantage that the stochastic terms in $\dd E_i$ vanish under time-averaging.
For a stochastic differential equation (SDE) of variable $X$(vector) with drift $\mu$(vector) and diffusion $\sigma$(matrix)
\begin{equation} \label{eqnS:SDE_general}
\dd X = \mu \dd t + \sigma \dd W ,
\end{equation}
where $\dd W$ is a vector consisting of standard Wiener processes,
Ito's formula gives the SDE of function $f(X)$
\begin{equation} \label{eqnS:SDE_ito}
\dd f(X) = ((\nabla_X^Tf)\mu + \frac{1}{2}\tr[\sigma \sigma^T \nabla_X\nabla_X^Tf])\dd t + (\nabla_X^Tf) \sigma \dd W,
\end{equation}
where $\nabla_X$ denotes the gradient with respect to $X$, the superscript $T$ denotes the transpose, and $\tr$ denotes the trace.

We begin by writing the equation of motion of our system \eqnname~(1) in the form of a stochastic differential equation \eqnname~\eqref{eqnS:SDE_general}.
We represent $N$ particles' position by a column vector $z = \sum_{i=1}^N \ket{i}\otimes z_i$, where $\ket{i}$ denotes the 2D subspace of particle $i$. Similar representations are also applied to $v$ and $\eta$. 
Then we get
\begin{align}
X &= \pmqty{ z & v & \eta }^T, \label{eqnS:SDE_X}\\
\mu &= \pmqty{ v \\
\frac{1}{m}(-\nabla_zU - BAv - \gamma v + \eta) \\
-\frac{1}{\tau}\eta }, \label{eqnS:SDE_mu} \\
\sigma &= \text{diag} \pmqty{0 & 0 & \frac{\sqrt{2\gamma T_a}}{\tau} I}, \label{eqnS:SDE_sigma}
\end{align}
where $U$ is the total energy of the system, $A$ is an antisymmetric matrix $A=\sum_i \ket{i}\bra{i}\otimes \mqty(0 & 1 \\ -1 & 0)$, and $\text{diag}()$ means a block-diagonal matrix.

Now we apply Ito's formula \eqnname~\eqref{eqnS:SDE_ito} to our system by associating with the function $f(X)$, the energy of particle $i$, $E_i(X)$.
The nonzero terms in the gradient of $E_i$ are
\begin{align}
\nabla_{z_i}E_i &= -(F_{ii} + \frac{1}{2}\sum_jF_{ji}), \\
\nabla_{z_j}E_i &= -\frac{1}{2}F_{ij}, \\
\nabla_{v_i}E_i &= m_iv_i.
\end{align}

The term $(\nabla_X^TE_i)\mu$ reads
\begin{equation}
\begin{split}
(\nabla_X^TE_i)\mu
&= (\nabla_{z_i}^TE_i)v_i + \sum_j(\nabla_{z_j}^TE_i)v_j + (\nabla_{v_i}^TE_i)m_i^{-1}(-\nabla_{z_i}U - BAv_i - \gamma v_i + \eta_i) \\
&= -(F_{ii} + \frac{1}{2}\sum_jF_{ji})^T v_i - \sum_j\frac{1}{2}F_{ij}^T v_j + v_i^T (F_{ii} + \sum_jF_{ji}) - \gamma v_i^Tv_i + v_i^T\eta_i \\
&= -\sum_j\frac{1}{2}(v_i + v_j)^T F_{ij} - \gamma v_i^Tv_i + v_i^T\eta_i ,
\end{split}
\end{equation}
where we used $F_{ji} = -F_{ij}$ and $v_i^TAv_i = 0$.
The term $\frac{1}{2}\tr[\sigma \sigma^T \nabla_X\nabla_X^Tf]$ and $\nabla_X^Tf$ are zero.

Finally, the energy change can be written as
\begin{align}
\dd E_i &= -\sum_jJ_{ij}\dd t + h_i \dd t, \label{eqnS:flux_dEi} \\
J_{ij} &= \frac{1}{2}(v_i + v_j)^T F_{ij}, \label{eqnS:flux_Jij} \\
h_i &= -\gamma_i v_i^Tv_i +v_i^T\eta_i. \label{eqnS:flux_hi}
\end{align}
$J_{ij}$ is identified as the energy transferred per unit time from particle $i$ to $j$, and $h_i$ is identified as the energy transferred from the bath to particle $i$.

As for the steady-state average of $J_{ij}$, we use $\expval{\dd U_{ij} / \dd t =0}$ and the chain rule to simplify \eqnname~\eqref{eqnS:flux_Jij}
\begin{equation}
    0 = v_i^T F_{ji} + v_j^T F_{ij}
    = -v_i^T F_{ij} + v_j^T F_{ij},
\end{equation}
and arrive at the expresion
\begin{equation}
    \expval{J_{ij}} = \expval{v_j^T F_{ij}}.
\end{equation}

\section{Numerical method for solving the energy flux}
One straightforward numerical method to compute the flux $J$ is as follows.
Our system is determined by the network geometry and parameters $m, k_g, k, B, \gamma, \tau, T_a$.
Given the equation of motion \eqnname~\eqref{eqnS:SDE_general},\eqref{eqnS:SDE_X}-\eqref{eqnS:SDE_sigma}, one can numerically solve for the covariance $C=\expval{XX^T}$ from the matrix equation $\mu C + C \mu^T = \sigma\sigma^T$ \cite{Gardiner2009ItoCalculus,Ceriotti2010ColoredNoiseThermostats}.
Finally, the flux \eqnname~\eqref{eqnS:flux_Jij}, which is bilinear in $x$ and $v$, can be extracted from the covariance $C$.
Numerical calculations of $\expval{J}$ are performed using Mathematica with custom code \cite{WolframResearch2018MathematicaVersion}

\section{Energy flux from linear response theory}
Following \cite{Kundu2011LargeDeviations}, we derive the expression of the energy flux (\eqnname~(7) and (10) in the main text) using a spectral linear response theory.

\subsection{Fourier modes for energy flux}

We define Fourier transform (FT) of a function $f(t)$ as
\begin{align}
\tilde{f}(\omega) &= \frac{1}{t} \int_0^t dt'\ f(t')e^{-i\omega t'},\quad
\omega = \frac{2\pi n}{t} ,\\
f(t) &= \sum_{\omega=-\infty}^{\infty} \tilde{f}(\omega) e^{i\omega t} .
\end{align}

The FT of the equation of motion \eqnname~\eqref{eqnS:SDE_general},\eqref{eqnS:SDE_X}-\eqref{eqnS:SDE_sigma} reads
\begin{align}
\tilde{v}(\omega) &= i\omega \tilde{z}(\omega) ,\label{eqnS:FT_v}\\
\tilde{z}(\omega) &= G^+(\omega) \tilde{\eta}(\omega) ,\label{eqnS:FT_z}\\
\tilde{\eta}(\omega) &= \frac{\sqrt{2\gamma T_a}}{1 + i\omega \tau} \tilde{\xi}(\omega) ,\label{eqnS:FT_eta}
\end{align}
where $G^+$ is the response function
\begin{equation} \label{eqnS:response}
G^{\pm}(\omega) = [K \pm i\omega(\gamma I + BA) - m\omega^2I]^{-1} .
\end{equation}

The energy flux $J_{ij}$ from \eqnname~\eqref{eqnS:flux_Jij} can be expressed as a bilinear function in $z$ and $v$, by writing the linearized force $F$ in terms of $z$,
\begin{align}
    J_{ij} &= kv^TA^{J}z \\
    \begin{split}
    A^{J} &\equiv \frac{1}{2} (\ket{i}\bra{i} \otimes e_{ij}e_{ij}^T + \ket{i}\bra{j} \otimes e_{ij}e_{ji}^T \\
    &\quad + \ket{j}\bra{i} \otimes (-e_{ji}e_{ij}^T) + \ket{j}\bra{j} \otimes (-e_{ji}e_{ji}^T)) .
    \end{split}
\end{align}
    
This bilinear form enables us to write the time integral of energy flux $Q = \int_0^t \dd{t'} J(t')$ as a sum of Fourier modes $\tilde{q}_\omega$ using Parseval's theorem,
\begin{align}
    Q &= t\sum_{\omega=-\infty}^{\infty} \tilde{q}_\omega, \\
    q_\omega &= k\tilde{v}^T A^{J} \tilde{z}^*
    = i\omega k \tilde{\eta}^TG^{+T}A^{J}G^-\tilde{\eta}^* ,
\end{align}
where the superscript $*$ denotes the complex conjugate.

Pairing $\tilde{q}_\omega$ and its conjugate $\tilde{q}_{-\omega}$ gives a real function, which would be beneficial for subsequent derivations.
\begin{gather}
    Q = t\sum_{\omega=2\pi/t}^{\infty} (\tilde{q}_\omega + \tilde{q}_{-\omega}), \label{eqnS:qmode_sum}\\
    \tilde{q}_\omega + \tilde{q}_{-\omega} = \tilde{\eta}(\omega)^T A_\omega^q \tilde{\eta}(\omega)^* \\
    A_\omega^q = -i\omega k G^{+T}(\omega) A^{as} G^-(\omega), \\
    A^{as} = -(A^{J} - {A^{J}}^T)
    = -\ket{i}\bra{j} \otimes e_{ij}e_{ji}^T + \ket{j}\bra{i} \otimes e_{ji}e_{ij}^T.
\end{gather}

Averaging $\tilde{q}_\omega + \tilde{q}_{-\omega}$ over the noise $\tilde{\eta}(\omega)$ using the relationship between $\tilde{\eta}$ and $\tilde{\xi}$ \eqnname~\eqref{eqnS:FT_eta}, and $\expval{\tilde{\xi}(\omega) \tilde{\xi}^T(\omega')} = \frac{1}{t} I \delta(\omega + \omega')$, we get
\begin{equation} \label{eqnS:qmode_single}
\begin{split}
\expval{\tilde{q}_\omega + \tilde{q}_{-\omega}}
&= \frac{2\gamma T_a}{1 + \omega^2 \tau^2} \tr[A_\omega^q \expval{\tilde{\xi}(-\omega)\tilde{\xi}(\omega)^T}] \\
&= \frac{1}{t} \frac{2\gamma T_a}{1 + \omega^2 \tau^2} \tr A_\omega^q .
\end{split}
\end{equation}

\subsection{Integrating over the Fourier modes}

In long time limit, the sum can be approximated by an integral
\begin{equation}
\frac{1}{t} \sum_{\omega=2\pi/t}^{\infty}
= \frac{1}{2t} \sum_{\omega=-\infty}^{\infty} \frac{t \Delta\omega}{2\pi}
\approx \frac{1}{4\pi}\int_{-\infty}^{\infty} \dd{\omega} .
\end{equation}
\eqnname~\eqref{eqnS:qmode_sum} and \eqref{eqnS:qmode_single} can then be turned to an integral expression of the flux
\begin{equation} \label{eqnS:flux_integral_raw}
\expval{J} = \lim_{t\rightarrow \infty} \frac{\expval{Q}}{t}
= \frac{\gamma T_a}{2\pi} \int_{-\infty}^{\infty} \dd{\omega} \frac{\tr A_\omega^q}{1+\omega^2\tau^2} .
\end{equation}

In the next steps, we simplify this integral with the help of the property \cite{Kundu2011LargeDeviations}
\begin{equation} \label{eqnS:response_property}
G^-(\omega) - G^{+T}(\omega) = 2i\omega\gamma G^-(\omega)G^{+T}(\omega) .
\end{equation}
Using this property, the trace of $A_\omega^q$ becomes
\begin{equation}
\begin{split}
\tr A_\omega^q &= -i\omega k \tr G^{+T} A^{as} G^- \\
&= -i\omega k \frac{1}{2i\omega \gamma} \tr (G^- - G^{+T})A^{as} \\
&= -\frac{k}{\gamma} \Re\tr G^+A^{as} .
\end{split}
\end{equation}

Plugging this trace into \eqnname~\eqref{eqnS:flux_integral_raw}, we get the integral form for the flux \eqnname~(7)
\begin{equation}
\expval{J} = -\frac{T_ak}{2\pi} \int_{-\infty}^{\infty} \dd{\omega} \frac{\Re \tr G^+A^{as}}{1+\omega^2\tau^2} .
\end{equation}

This integral can be calculated using the residue theorem.
Since $\Im G^+(-\omega) = -\Im G^+(\omega)$,
$\frac{\Im \tr G^+A^{as}}{1+\omega^t\tau^2}$ is an odd function of
$\omega$, and its line integral vanishes.
\begin{equation}
\expval{J} = -\frac{T_ak}{2\pi} \int_{-\infty}^{\infty} \dd{\omega} \frac{\tr G^+A^{as}}{1+\omega^2\tau^2} .
\end{equation}
The integrand vanishes at $\omega \rightarrow \infty$, so the line
integral can be converted to a contour integral along the
counter-clockwise semicircle $R$ in the lower-half plane
\begin{equation}
\expval{J} = \frac{T_ak}{2\pi} \oint_R \dd{\omega} \frac{\tr G^+A^{as}}{1+\omega^2\tau^2} .
\end{equation}
The noise correlation $\tau$ introduces a pole of the integrand at $\omega=-i/\tau$, thus the contour integral can be evaluated as
\begin{equation} \label{eqnS:flux_residue}
\expval{J} = -\frac{T_ak}{2\tau} \tr G^+(-\frac{i}{\tau})A^{J,as} ,
\end{equation}
and the response function at $\omega=-i/\tau$ reads
\begin{equation} \label{eqnS:response_at_pole}
G^+(-\frac{i}{\tau})
= [K + (\frac{\gamma}{\tau} + \frac{m}{\tau^2})I + \frac{B}{\tau}A]^{-1} .
\end{equation}

In theory, the equation \eqnname~\eqref{eqnS:flux_residue} provides the analytical solution of the flux, because the inverse matrix \eqnname~\eqref{eqnS:response_at_pole} can be expressed analytically.
In practice, analytical solutions can be easily calculated for small networks, but are hard for large networks.
Nevertheless, some general properties of the flux can be obtained from \eqnname~\eqref{eqnS:flux_residue} after some algebra. For network with only horizontal and vertical bonds (\figurename~1b), all fluxes are zero. For two networks whose slanted bonds have opposite angles (\figurename~1b), their fluxes are opposite. Changing $B$ to $-B$ would change the flux $J$ to $-J$.

\section{Kirchoff's law}
The derivation of the Kirchoff's law is similar to the derivation of the energy flux, except that we use the energy from the bath to the particle $h_i$ in \eqnname~\eqref{eqnS:flux_hi} instead of $J_{ij}$ in \eqnname~\eqref{eqnS:flux_Jij}.

Following the procedure in the last section from \eqnname~\eqref{eqnS:qmode_sum} to \eqref{eqnS:flux_integral_raw}, we arrive at an integral expression for the $\expval{h_i}$ with a different $A_\omega^q$
\begin{gather}
    \expval{h_i} = \frac{\gamma T_a}{2\pi} \int_{-\infty}^{\infty} \dd{\omega} \frac{\tr A_\omega^q}{1+\omega^2\tau^2}, \label{eqnS:flux_hi_fourier}\\
    A_\omega^q = i\omega (G^{+T} \rho_i - \rho_iG^-) - 2\gamma\omega^2 G^{+T} \rho_i G^- ,\\
    \rho_i = \ket{i}\bra{i} .
\end{gather}

Using the property of $G^\pm$ \eqnname~\eqref{eqnS:response_property}, we get
\begin{equation}
    \begin{split}
        \tr (G^{+T} \rho_i - \rho_iG^-)
        &= \tr \rho_i(G^{+T} - G^-) \\
        &= -\tr \rho_i 2i\omega \gamma G^- G^{+T} \\
        &= -2i\omega\gamma \tr G^{+T}\rho_i G^- ,
    \end{split}
\end{equation}
so the trace of $A_\omega^q$ vanishes
\begin{equation}
    \tr A_\omega^q
    = i\omega \tr (G^{+T} \rho_i - \rho_iG^-) - \tr 2\gamma\omega^2 G^{+T} \rho_i G^-
    = 0 .
\end{equation}

From \eqnname~\eqref{eqnS:flux_hi_fourier}, $\expval{h_i}$ is also zero, so on average there is no energy exchange between the particle and the bath. Because the average change of $E_i$ is zero, and $\expval{\dot{E_i}} = -\sum_j\expval{J_{ij}} + \expval{h_i}$, we obtain the Kirchoff's law
\begin{equation}
    -\sum_j\expval{J_{ij}} = \sum_j\expval{J_{ji}} = 0 .
\end{equation}

\section{Connection to isolated gyroscopic networks}

\begin{figure}[tbp]
	\centering
	\includegraphics[width=0.8\textwidth]{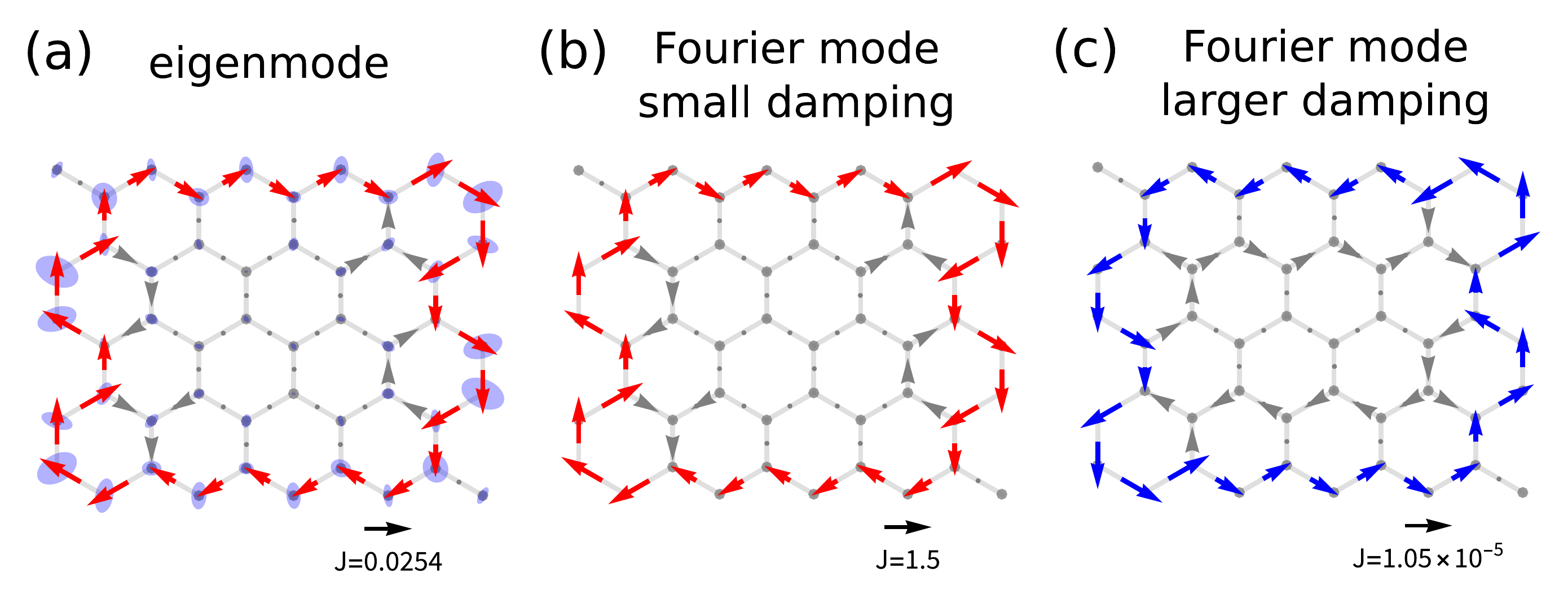}
    \caption{
        Comparison between a boundary-localized eigenmode of the undamped isolated network and the Fourier modes of the damped network at the same frequency. First order dynamics (by setting $m=0$) are used. Numerical calculations are performed with all other parameters set to $1$.
        (a) Eigenmode of undamped gyroscopic system. For the frequency chosen, the eigenmode is localized on the boundary. Blue disks represent the orbit of particles.
        (b) The Fourier mode of damped variant of our model at small $\gamma$ ($\gamma=0.001$) resembles the eigenmode.
        (c) The Fourier mode at larger $\gamma$ ($\gamma=1$) is no longer close to the eigenmode.
    }
    \label{fig:Fourier_modes}
\end{figure}

Since our model is built upon the well-studied isolated system in Refs \cite{Nash2015TopologicalMechanics,Susstrunk2016ClassificationTopological,Mitchell2018AmorphousTopological,Lee2018TopologicalDynamics}, we would like to build a connection between our energy flux in the active system and eigenmodes in those studies.
In this section, we show that the flux formula \eqnname~(10) can be decomposed to a weighted sum over eigenmodes \eqnname~\eqref{eqnS:flux_eigen}. Then we apply this result to a honeycomb network as an example.

The Fourier analysis from \secname~IV in the main text is not suitable for this connection, because Fourier modes and eigenmodes are related only at small $\gamma$'s (\figurename~\ref{fig:Fourier_modes}a and b), but they become dissimilar at larger $\gamma$'s (\figurename~\ref{fig:Fourier_modes}a and c).
The underlying discrepancy between Fourier modes and eigenmodes comes from the fact that eigenmodes are a natural basis for the isolated network, whereas Fourier modes have an extra factor of friction or damping.
In addition to this extra factor $\gamma$,
the active system also has extra factors of $m$ and $\tau$. The factor $m$ comes from the order of dynamics: the active system is second order in time, while the gyroscopic dynamics in \cite{Nash2015TopologicalMechanics} is first order, which corresponds to the $m\rightarrow 0$ limit.

Our starting point is \eqnname~(10). The key point is that, in the function $G^+(-i/\tau)$ from the equation, $\gamma,m,\tau$ are not independent factors. Rather, they act collectively through 
\begin{equation}
    k_{g,\tau} \equiv k_g+\frac{\gamma}{\tau}+\frac{m}{\tau^2}.
\end{equation}
In effect, the extra factors $m,\gamma, \tau$ only add a modification to $k_g$.
Following these ideas, we imagine a reference isolated system with modified on-site spring constant $k_{g,\tau}$. Then after some algebra, the flux $\expval{J}$ in active system can be written as a weighted sum of the flux of each eigenmode $J^{\text{eig}}_{\omega_e}$ in the reference system (see the next section for the derivation),
\begin{equation} \label{eqnS:flux_eigen}
    \expval{J} = \sum_{\omega_e} \frac{1}{1+\omega_e^2\tau^2} J^{\text{eig}}_{\omega_e}.
\end{equation}
Here $\omega_e$ is the discrete eigen-frequency of the reference system, not to be confused with the continuous Fourier frequency $\omega$. The amplitude of eigenmode is set such that its energy is $T_a$, and $J^{\text{eig}}_{\omega_e}$ is the time-averaged energy flux.

A related result is a so called ``sum rule", namely, the unweighted sum of all modes is zero, 
\begin{align}
    \sum_{\omega_e} J^{\text{eig}}_{\omega_e} = 0. \label{eqnS:eigen_sum_rule}
\end{align}
This ``sum rule" can be derived from direct calculations (see the next section).

From this eigenmode decomposition, the discussion of time-reversal symmetry in the isolated system \cite{Nash2015TopologicalMechanics} immediately carries over to the active system. For network geometries that satisfy time-reversal symmetries, the energy flux of eigenmodes are zero. Thus through \eqnname~\eqref{eqnS:flux_eigen}, the flux in active system is also zero. This result can alternatively be obtained from \eqnname~(10) through some linear algebra.

\begin{figure}[tbp]
	\centering
	\includegraphics[width=0.8\textwidth]{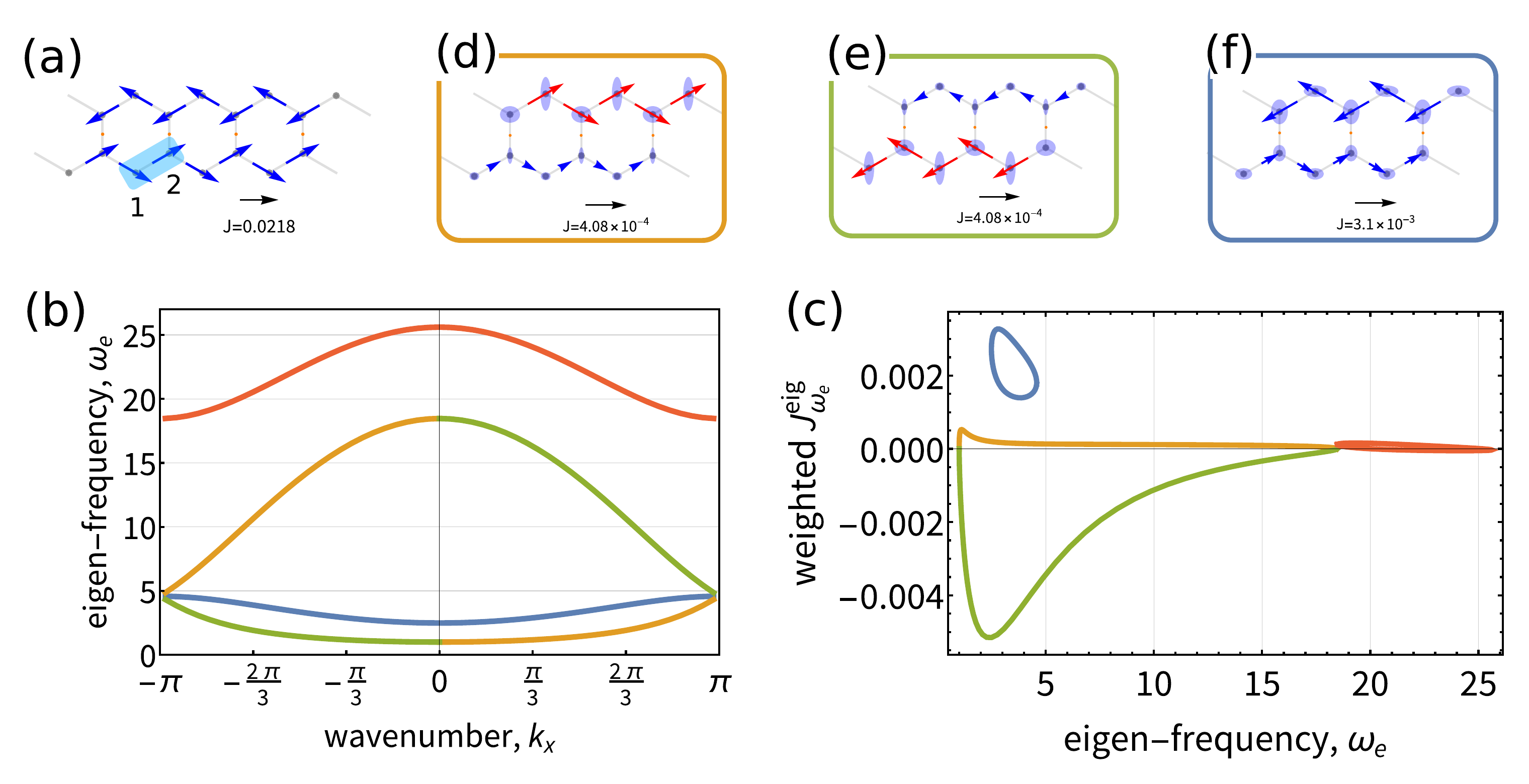}
    \caption{
        Using the eigenmode decomposition, we explain how the flux in honeycomb network is CCW, even though its edgemodes contribute to CW fluxes.
        (a) Network used for calculation, which consists of one row of hexagons (51 unit cells) and has periodic boundary in $x$ direction. Parameters: $k_{g,\tau}=1, k=10$, others are $1$.
        (b) Band structure of the network (marked with different colors). The yellow/green band contains CW flux localized on the top/bottom edge (an example mode is shown in (d)/(e)). The blue band contains bulk modes with CCW flux (also see (f)).
        (c) Weighted flux $J_{\omega_e}^{eig}$ from $1$ to $2$ (marked in (a)) of the four bands. Total flux in the green band with CW edge modes and the blue band with CCW bulk modes are $-0.106$ and $0.115$, respectively. As a result, the net flux is CCW.
    }
    \label{fig:eigen_modes}
\end{figure}

As an application, we will analyze the flux in the honeycomb network using the eigenmode decomposition \eqnname~\eqref{eqnS:flux_eigen} and the ``sum rule" \eqnname~\eqref{eqnS:eigen_sum_rule}.
The flux pattern in the active honeycomb network displays CCW flux localized on the boundary (\figurename~1b). This localization is reminiscent of the edgemode in \cite{Nash2015TopologicalMechanics} (\figurename~\ref{fig:Fourier_modes}b), however, their directions are opposite.
From the decomposition \eqnname~\eqref{eqnS:flux_eigen}, the edgemodes should contribute a large CW flux in the active system, but somewhat surprisingly, the net flux is CCW.
To better analyze the contribution from each eigenmode, we look at a simple honeycomb lattice with only one layer (\figurename~\ref{fig:eigen_modes}a).
This lattice has four bands (\figurename~\ref{fig:eigen_modes}b), two bulk bands (blue, red) and two edge bands (green, yellow). The weighted flux of each band is plotted in \figurename~\ref{fig:eigen_modes}c. We see that the CW edge band does contribute a large CW flux (green curve in \figurename~\ref{fig:eigen_modes}c), however, due to the ``sum rule", the unweighted sum of other bands has to be CCW. In the honeycomb lattice, it happens that many of this CCW fluxes are contained in the lower bulk band (blue curve in \figurename~\ref{fig:eigen_modes}c and example mode in \figurename~\ref{fig:eigen_modes}f). When the flux gets weighted, the CCW flux from lower bulk band outweighs CW flux from the edgemodes, the other two bands (yellow and red curve in \figurename~\ref{fig:eigen_modes}c) also contribute to CCW flux, although relatively small. As a result, the net flux is CCW, which is opposite to the flux of the edgemode.

\section{Derivation of eigenmode decomposition of energy flux}

To derive the eigenmode decomposition \eqnname~\eqref{eqnS:flux_eigen}, we first look at the reference isolated system, write down its eigenmodes \eqnname~\eqref{eqnS:mode_zct} and time-averaged energy flux \eqnname~\eqref{eqnS:mode_J_isolated}. Then we turn to the active system and decompose the flux \eqnname~\eqref{eqnS:flux_residue} using the eigenmodes to get \eqnname~\eqref{eqnS:mode_J_active}. Finally we show that the flux from these two sides are actually related in \eqnname~\eqref{eqnS:mode_J_relation}.
Lastly we also derive the ``sum rule" \eqnname~\eqref{eqnS:mode_J_sum}.

\subsection{Reference isolated system}
The reference isolated system has first-order gyroscopic dynamics as in \cite{Nash2015TopologicalMechanics}. In the setup with Lorentz force, the dynamical equation can be obtained by setting the mass to zero, and replacing the force matrix $K$ by ${K^\tau \equiv K + (\frac{\gamma}{\tau} + \frac{m}{\tau^2})I}$
\begin{equation}
    \dot{z} = \frac{1}{B} A K^\tau z.
\end{equation}
Following \cite{Nash2015TopologicalMechanics}, we convert to complex representation with $z^c \equiv \pmqty{x + iy & x - iy}^T$
\begin{equation}
i \dot{z}^c = \Omega z^c,\quad
K^\tau = i B A O^{-1} \Omega O
\end{equation}
where $O,O^{-1}$ are the transformations between $z$ and $z^c$ $z^c = Oz, z = O^{-1}z^c$.

Writing the eigenvalue problem as
\begin{equation}
\Omega u_{\omega_e} = \omega_e u_{\omega_e} ,
\end{equation}
the eigenmode with eigen-frequency $\omega_e$ reads
\begin{equation} \label{eqnS:mode_zct}
z^c_{\omega_e}(t) =  (u_{\omega_e} e^{-i\omega_e t} + u_{-\omega_e} e^{i\omega_e t})z_0 ,
\end{equation}
where $z_0$ is the amplitude, and it will be specified shortly.
The eigenmode needs a combination of $\omega_e$ and $-\omega_e$ to ensure that the motion of $x$ and $y$ is real-valued.
Mathematically, this combination is possible because of a symmetry in this eigenvalue problem, when there is $\omega_e$, there is also solution $-\omega_e$ with $u_{-\omega_e} = \pmqty{ 0 & I \\ I & 0 } u_{\omega_e}^*$.

A related property we need later is that, the left eigenvector $v_{\omega_e}$ can be expressed as $v_{\omega_e} = c_{\omega_e} \pmqty{ -I & 0 \\ 0 & I } u_{\omega_e}$,
where $c_{\omega_e}$ is a real prefactor to ensure normalization $v_{\omega_e}^T u_{\omega_e} = 1$.
If there are degenerate eigenvectors (like $v_{\omega_e}^{1},v_{\omega_e}^{2},\dots$), we choose an orthonormal basis set, i.e. $v_{\omega_e}^{i,T} u_{\omega_e}^{j} = 0$ for $i \neq j$.
With the introduction of $c_{\omega_e}$, we now set the amplitude $z_0$ to
$z_0^2 = -2 c_{\omega_e} T_a / \omega_e B$, such that the energy of the eigenmode is $T_a$.

The instantaneous energy flux $J_{\omega_e}$ of mode $z^c_{\omega_e}$ writes
\begin{equation}
\begin{split}
J_{\omega_e} &= (O^{-1} v^{c}_{\omega_e})^T A^J O^{-1}z^c_{\omega_e} \\
&= \tr O^{-1,T} A^J O^{-1}z^c_{\omega_e} v^{cT}_{\omega_e} .
\end{split}
\end{equation}
From the expression of mode \eqnname~\eqref{eqnS:mode_zct},
\begin{equation}
z^c_{\omega_e} v^{cT}_{\omega_e} = -i\omega_e(u_{\omega_e} e^{-i\omega_e t} + u_{-\omega_e}e^{i\omega_e t})(u_{\omega_e}^T e^{-i\omega_e t} - u_{-\omega_e}^Te^{i\omega_e t})z_0^2 .
\end{equation}
When averaging over time, terms like $e^{\pm 2i\omega_e t}$ vanish, so we get
\begin{equation}
\overline{z^c_{\omega_e} v^{cT}_{\omega_e}} = i\omega_e(u_{\omega_e} u_{-\omega_e}^T - u_{-\omega_e} u_{\omega_e}^T)z_0^2 .
\end{equation}
Plugging in $z_0^2 = -2 c_{\omega_e} T_a / \omega_e B$, the time-averaged flux of the eigenmode $J_{\omega_e}^\text{eig}$ reads
\begin{equation} \label{eqnS:mode_J_isolated}
J_{\omega_e}^\text{eig} = -\frac{2T_ak}{B} ic_{\omega_e} \tr O^{-1,T}A^JO^{-1} (u_{\omega_e} u_{-\omega_e}^T - u_{-\omega_e} u_{\omega_e}^T) .
\end{equation}

\subsection{Active system}
Now we turn to the active system, and the starting point is \eqnname~\eqref{eqnS:flux_residue}.
We need to decompose $G^\tau \equiv G^+(-i/\tau)$ into modes as below,
\begin{align}
G^\tau &= \frac{i}{B} O^{-1} (\Omega - \frac{i}{\tau}I)^{-1} OA ,\\
(\Omega - \frac{i}{\tau}I)^{-1} &=
\sum_{\omega_e > 0}\frac{i\tau}{1 + \omega_e^2 \tau^2}  (u_{\omega_e} v_{\omega_e}^T + u_{-\omega_e} v_{-\omega_e}^T) + \\
&\quad \sum_{\omega_e > 0}\frac{\omega_e\tau^2}{1 + \omega_e^2 \tau^2}  (u_{\omega_e} v_{\omega_e}^T - u_{-\omega_e} v_{-\omega_e}^T) ,\\
G^\tau &= \sum_{\omega_e > 0}\frac{-\tau/B}{1 + \omega_e^2 \tau^2} O^{-1} (u_{\omega_e} v_{\omega_e}^T + u_{-\omega_e} v_{-\omega_e}^T) OA + \\
&\quad \sum_{\omega_e > 0}\frac{i\omega_e\tau^2/B}{1 + \omega_e^2 \tau^2} O^{-1} (u_{\omega_e} v_{\omega_e}^T - u_{-\omega_e} v_{-\omega_e}^T) OA .
\end{align}
The averaged flux $\expval{J}$ reads
\begin{equation}
\begin{split}
\expval{J} &=
\sum_{\omega_e > 0}\frac{T_ak/(2B)}{1 + \omega_e^2 \tau^2} \tr O^{-1} (u_{\omega_e} v_{\omega_e}^T + u_{-\omega_e} v_{-\omega_e}^T) OAA^{as} + \\
&\quad \sum_{\omega_e > 0}\frac{-i\omega_e T_ak\tau/(2B)}{1 + \omega_e^2 \tau^2} \tr O^{-1} (u_{\omega_e} v_{\omega_e}^T - u_{-\omega_e} v_{-\omega_e}^T) OAA^{as}.
\end{split}
\end{equation}
The second term can be shown to be zero,
$\tr OAA^{as}O^{-1} (u_{\omega_e} v_{\omega_e}^T - u_{-\omega_e} v_{-\omega_e}^T) = 0$.
So the mode decomposition in its preliminary form reads
\begin{gather}
\expval{J} = \sum_{\omega_e} \expval{J}_{\omega_e} ,\\
\expval{J}_{\omega_e} \equiv \frac{T_ak}{2B}\frac{1}{1 + \omega_e^2 \tau^2} \tr OAA^{as}O^{-1} (u_{\omega_e} v_{\omega_e}^T + u_{-\omega_e} v_{-\omega_e}^T). \label{eqnS:mode_J_active}
\end{gather}

\subsection{The relationship between isolated system and active system}
Now we need to find relationship between these two fluxes $\expval{J}_{\omega_e}$ \eqnname~\eqref{eqnS:mode_J_active} and $J_{\omega_e}^\text{eig}$ \eqnname~\eqref{eqnS:mode_J_isolated}.
We will write $J_{\omega_e}^\text{eig}$ in a form that looks similar to
$\langle J\rangle_{\omega_e}$.
Converting $A^{J}$ to $A^{as}$ using $A^{as}=-(A_J-A_J^T)$, and $u_{\omega_e}$ to $v_{\omega_e}$ using $v_{\omega_e} = c_{\omega_e} \pmqty{ -I & 0 \\ 0 & I } u_{\omega_e}$, we get
\begin{equation}
J_{\omega_e}^\text{eig} = -\frac{iT_ak}{B} \tr AO^{-1,T}A^{as}O^{-1} (u_{\omega_e} v_{\omega_e}^T + u_{-{\omega_e}}v_{-{\omega_e}}^T) .
\end{equation}
From direct calculation, $AO^{-1,T} = \frac{i}{2} OA$, and $J_{\omega_e}^\text{eig}$ becomes the same as $\expval{J}_{\omega_e}$ apart from a factor
\begin{equation} \label{eqnS:mode_J_isolated_mod}
J_{\omega_e}^\text{eig} = \frac{T_ak}{2B} \tr OAA^{as}O^{-1} (u_{\omega_e} v_{\omega_e}^T + u_{-{\omega_e}}v_{-{\omega_e}}^T) .
\end{equation}

Comparing \eqnname~\eqref{eqnS:mode_J_isolated_mod} with \eqref{eqnS:mode_J_active}, we derived the relationship between flux from active system and isolated system
\begin{equation} \label{eqnS:mode_J_relation}
\expval{J}_{\omega_e} = \frac{1}{1 + {\omega_e}^2 \tau^2} J_{\omega_e}^\text{eig} .
\end{equation}

Lastly, we show that the unweighted sum of $J_{\omega_e}^\text{eig}$ is zero.
This unweighted sum reads
\begin{equation}
\sum_{\omega_e} J_{\omega_e}^\text{eig} = \frac{T_ak}{2B} \tr [O A A^{as} O^{-1} UV^T],
\end{equation}
where $U$ is the collection of all right eigenvectors $U = \pmqty{u_{\omega_e,1} & u_{\omega_e,2} & \cdots}$, and likewise for $V$.
Since $UV^T = I$ from orthonormality, this sum vanishes
\begin{equation} \label{eqnS:mode_J_sum}
\sum_{\omega_e} J_{\omega_e}^\text{eig} = \frac{T_ak}{2B} \tr A A^{as} = 0.
\end{equation}

\section{Path summation of energy flux}
To derive the path summation formula \eqnname~(11) and the path rules, we start from \eqnname~\eqref{eqnS:flux_residue}, expand around a noninteracting reference system with $k=0$ to get \eqnname~\eqref{eqnS:smallk_expand}, discuss the convergence radius in \eqnname~\eqref{eqnS:smallk_convergence}, then insert resolution of identity to make each term representable by a path as in \eqnname~\eqref{eqnS:smallk_Sl}, and arrive at the path summation formula in \eqref{eqnS:smallk_path_sum}.
We also provide a convenient way to calculate $S_{-l}$ in \eqnname~\eqref{eqnS:smallk_S-l}, and a heuristic interpretation of $S_l$ in \eqnname~\eqref{eqnS:smallk_path_vector}.

\subsection{Derivation of path summation formula}
Similar to the last section, the central object is $G^\tau$.
In the noninteracting case ($k=0$), $G^\tau$ is analytically solvable. We denote $G^\tau(k=0) = G^\tau_0$.
The inverse $(G^\tau_0)^{-1}$ has a block diagonal form,
\begin{equation}
    (G^\tau_0)^{-1} = k_{g,\tau} I + \frac{B}{\tau}A
    = \sum_i \ketbra{i}{i} \otimes (k_{g,\tau}I + \frac{B}{\tau} \pmqty{0 & 1 \\ -1 & 0}).
\end{equation}
where $k_{g,\tau} \equiv k_g + \frac{\gamma}{\tau} + \frac{m}{\tau^2}$.
Then $G^\tau_0$ is also block diagonal, with each block the inverse of the blocks above,
\begin{equation} \label{eqnS:smallk_Gtau0}
    G^\tau_0 = \sum_i \ketbra{i}{i} \otimes \frac{1}{(k_{g,\tau})^2 + (B/\tau)^2}(k_{g,\tau} I - \frac{B}{\tau} \pmqty{0 & 1 \\ -1 & 0})
    = \sum_i \ketbra{i}{i} \otimes \frac{1}{k_0}R_\alpha,
\end{equation}
where $k_0 \equiv \sqrt{(k_{g,\tau})^2 + (B/\tau)^2}$, and $R_\alpha$ is the rotation matrix with angle $\alpha \equiv \arcsin{\frac{B/\tau}{k_0}}$, $R_\alpha = \pmqty{\cos\alpha & -\sin\alpha \\ \sin\alpha & \cos\alpha}$.

We now turn on $k$. We denote the inter-particle part of the force matrix $K$ as $k K_s$, where the factor $k$ is extracted so that the matrix $K_s$ is dimensionless. The blocks of $K_s$ read
\begin{equation} \label{eqnS:smallk_matKs}
    \mel{i}{K_s}{i} = \sum_{i'} e_{ii'}e_{ii'}^T, \quad
    \mel{i}{K_s}{j} = e_{ij}e_{ji}^T.
\end{equation}
Then $G^\tau$ reads
\begin{equation}
    G^\tau = \frac{1}{(G^\tau_0)^{-1} + k K_s}
    = \frac{1}{k_0} [(k_0 G^\tau_0)^{-1} + \frac{k}{k_0}K_s]^{-1}
\end{equation}
In small $k/k_0$ regime, this matrix inversion can be expanded as
\begin{equation} \label{eqnS:smallk_expand}
    \begin{split}
    G^\tau &= \frac{1}{k_0} [(k_0 G^\tau_0) + \frac{k}{k_0} (k_0 G^\tau_0) (-K_s) (k_0 G^\tau_0) + (\frac{k}{k_0})^2 (k_0 G^\tau_0) (-K_s) (k_0 G^\tau_0) (-K_s) (k_0 G^\tau_0) + \dots ]\\
    &= \frac{1}{k_0} (k_0 G^\tau_0) \sum_{n=0}^{\infty}[\frac{k}{k_0}(-K_s)(k_0 G^\tau_0)]^n.
    \end{split}
\end{equation}
To find the convergence radius, we can write the eigen-decomposition of the matrix $(-K_s)(k_0 G^\tau_0)$ as $(-K_s)(k_0 G^\tau_0) = W\Lambda W^{-1}$, where $\Lambda$ is the diagonal matrix that contains all eigenvalues $\lambda_i$'s, then the flux becomes
\begin{equation}
    \begin{split}
    \expval{J} &\propto \tr G^\tau A^{as}
    \propto \sum_{n=0}^{\infty}\tr (k_0 G^\tau_0) [\frac{k}{k_0}W\Lambda W^{-1}]^n A^{as} \\
    &= \sum_i [W^{-1} A^{as} (k_0 G^\tau_0) W]_{ii} \sum_n (\frac{k}{k_0} \lambda_i)^n.
    \end{split}
\end{equation}
For terms in the series to be convergent, $\frac{k}{k_0}$ should satisfy
\begin{equation} \label{eqnS:smallk_convergence}
    \frac{k}{k_0} < \frac{1}{\max_i\abs{\lambda_i}}.
\end{equation}

Before inserting resolution of identity to make paths, we note that the matrix $A^{as}$ and $K_s$ have common blocks, $A^{as} = -\ket{i}\bra{j} \otimes e_{ij}e_{ji}^T + \ket{j}\bra{i} \otimes e_{ji}e_{ij}^T$ and $\mel{i}{K_s}{j} = e_{ij}e_{ji}^T$, so that $A^{as}$ can merge with the series of $G^\tau$.
\begin{equation} \label{eqnS:smallk_J_block}
    \begin{split}
    \frac{\expval{J}}{T_a/\tau} = -\frac{k}{2} (\tr G^\tau A^{as})
    &= -\frac{k}{2} (\tr \mel{i}{G^\tau}{j} e_{ji}e_{ij}^T - \tr \mel{j}{G^\tau}{i} e_{ij}e_{ji}^T) \\
    &= \frac{k}{2} (\tr \mel{i}{G^\tau}{j} \mel{j}{-K_s}{i} - \tr \mel{j}{G^\tau}{i} \mel{i}{-K_s}{j}).
    \end{split}
\end{equation}

Now we use the expansion \eqnname~\eqref{eqnS:smallk_expand}, and look at the contribution of its $(n-1)$'th-order term to the first term of the flux, $k \tr \mel{i}{\frac{1}{k_0} (k_0 G^\tau_0) [\frac{k}{k_0}(-K_s)(k_0 G^\tau_0)]^{n-1}}{j} \mel{j}{-K_s}{i}$.
If $n-1=0$, this term vanishes, so we only need to consider $n-1 \ge 1$ case.
Insert $n-1$ resolution of identity $I = \sum_{l_a=1}^N \ketbra{l_a}{l_a}$, and plug in $k_0 G^\tau_0$ \eqnname~\eqref{eqnS:smallk_Gtau0} and $K_s$ \eqnname~\eqref{eqnS:smallk_matKs}, we get
\begin{equation}
    \begin{split}
    &\frac{k}{k_0} \tr \mel{i}{(k_0 G^\tau_0) [\frac{k}{k_0}(-K_s)(k_0 G^\tau_0)]^{n-1}}{j} \mel{j}{-K_s}{i} \\
    =& (\frac{k}{k_0})^n \sum_{l_1,l_2,\dots,l_{n-1}} \tr \bra{i} (k_0G^\tau_0) \ket{l_{n-1}}\bra{l_{n-1}} (-K_s) (k_0G^\tau_0) \cdots \ket{l_1}\bra{l_1} (-K_s) (k_0G^\tau_0) \ket{j} \mel{j}{-K_s}{i}\\
    =& (\frac{k}{k_0})^n \sum_{l_1,l_2,\dots,l_{n-2}} \tr R_\alpha (-K_s)_{il_{n-2}} R_\alpha \cdots (-K_s)_{l_1j} R_\alpha (-K_s)_{ji},
    \end{split}
\end{equation}
where $(-K_s)_{l_bl_a} \equiv \mel{l_b}{-K_s}{l_a}$.
We will denote path $l = i\rightarrow j\rightarrow l_1\rightarrow l_2\rightarrow \dots \rightarrow l_{n-2}\rightarrow i$, and its corresponding term in the above summation as $S_l$
\begin{equation} \label{eqnS:smallk_Sl}
    S_l = (\frac{k}{k_0})^n \tr R_\alpha (-K_s)_{il_{n-2}} R_\alpha \cdots (-K_s)_{l_1j} R_\alpha (-K_s)_{ji}.
\end{equation}

The second term of the flux in \eqref{eqnS:smallk_J_block} can be treated similarly, and it results in $S_{-l}$, where $-l$ means path $l$ in its reversed order.
Combining \eqnname~\eqref{eqnS:smallk_Sl} and \eqref{eqnS:smallk_J_block}, we get the path summation formula of the flux
\begin{equation} \label{eqnS:smallk_path_sum}
    \frac{\expval{J}}{T_a/\tau} = \sum_l J^\text{path}_l
    = \sum_l \frac{1}{2}(S_l - S_{-l}).
\end{equation}

\subsection{Path rules and discussions}
The path rules can be extracted from the expression of $S_l$ and $J^\text{path}$.
From the element $(-K)_{l_bl_a}$ in $S_l$, we see that either $l_a,l_b$ are bonded, or $l_a=l_b$, otherwise $(-K)_{l_bl_a}=0$.
So the path has to be a closed walk along the edges of the network.
From $J^\text{path}_l$ for flux from $i$ to $j$, we see that if the path contains equal numbers of $i\rightarrow j$ and $j\rightarrow i$, the net contribution is zero. Because, either $l=-l$, so $J^\text{path}_l \propto S_l - S_{-l} = 0$, or $l'\equiv -l$ is another path, and $J^\text{path}_l + J^\text{path}_{l'} = 0$.

To calculate $S_{-l}$, there is a convenient way given that $S_l$ is known.
Based on the transformation below, $S_{-l}$ can be obtained by taking the result of $S_l$, then replacing $\alpha$ by $-\alpha$.
\begin{equation} \label{eqnS:smallk_S-l}
    \begin{split}
    S_{-l} / (\frac{k}{k_0})^n
    &= \tr (R_\alpha (-K_s)_{ij} R_\alpha (-K_s)_{jl_1} \cdots R_\alpha  (-K_s)_{l_{n-2}i})^T \\
    &= \tr (-K_s)_{l_{n-2}i}^T R_\alpha^T \cdots (-K_s)_{jl_1}^T R_\alpha^T (-K_s)_{ij}^T R_\alpha^T \\
    &= \tr R_{-\alpha} (-K_s)_{il_{n-2}} R_{-\alpha} \cdots (-K_s)_{l_1j} R_{-\alpha} (-K_s)_{ji}.
    \end{split}
\end{equation}

To interpret $S_l$ in a more heuristic way, we insert $I = e_{ij}e_{ij}^T + e_{ij,\perp}e_{ij,\perp}^T$ to the trace in \eqnname~\eqref{eqnS:smallk_Sl}, where $e_{ij,\perp}$ denotes the unit direction perpendicular to $e_{ij}$.
Because $(-K_s)_{ji}e_{ij,\perp} = 0$, the trace reduces to a matrix product
\begin{equation} \label{eqnS:smallk_path_vector}
    S_l/(\frac{k}{k_0})^n = e_{ij}^T R_\alpha (-K_s)_{i l_{n-2}} R_\alpha \cdots (-K_s)_{l_1j} R_\alpha (-K_s)_{ji} e_{ij}.
\end{equation}
This expression means the following operations:
starting from a unit displacement of $i$ along $e_{ij}$, $j$ would be displaced according to the force $(-K_s)_{ji} e_{ij}$, after which $j$ is rotated by angle $\alpha$; then start from $j$ and perform similar operations for $(-K_s)_{l_1j}$ and $R_\alpha$; finally, the transmission goes back to $i$; we project the displacement onto $e_{ij}$, and this value is $S_l$ (apart from the prefactor $(\frac{k}{k_0})^n$).

\subsection{Flux of polygon paths}
Here we write down the flux formula for a polygon path without loops.
It is easier to work in local coordinates, where each node has its own coordinate system.
For node $i$ in the path, let the outer angle from $i$ to $i-1$ be $\pi$, and the angle from $i$ to $i+1$ be $\theta_i$.
Then the matrix $(-K_s)_{i+1,i}$ reads
\begin{equation}
    (-K_s)_{i+1,i}
    = -e_{i+1,i} e_{i,i+1}^T
    = -\pmqty{-1 \\ 0} \pmqty{\cos\theta_i & \sin\theta_i}
    = \pmqty{1 \\ 0} \pmqty{\cos\theta_i & \sin\theta_i}.
\end{equation}
The trace in $S_l$ becomes
\begin{equation}
    \begin{split}
    S_l / (\frac{k}{k_0})^n
    &= \tr \prod_i (-K_s)_{i+1,i} R_\alpha
    = \tr \prod_i \pmqty{1 \\ 0} \pmqty{\cos\theta_i & \sin\theta_i} R_\alpha \\
    &= \prod_i \pmqty{\cos\theta_i & \sin\theta_i} R_\alpha \pmqty{1 \\ 0}
    = \prod_i \cos(\theta_i - \alpha)
    \end{split}
\end{equation}
So the flux for this path without loops writes
\begin{equation} \label{eqnS:smallk_path_polygon}
    J^\text{path}_\text{polygon} = \frac{1}{2} (\frac{k}{k_0})^n (\prod_i \cos(\theta_i - \alpha) - \prod_i \cos(\theta_i + \alpha)).
\end{equation}

\subsection{Contribution from higher-order paths}
\begin{figure}[tbp]
	\centering
	\includegraphics[width=0.5\textwidth]{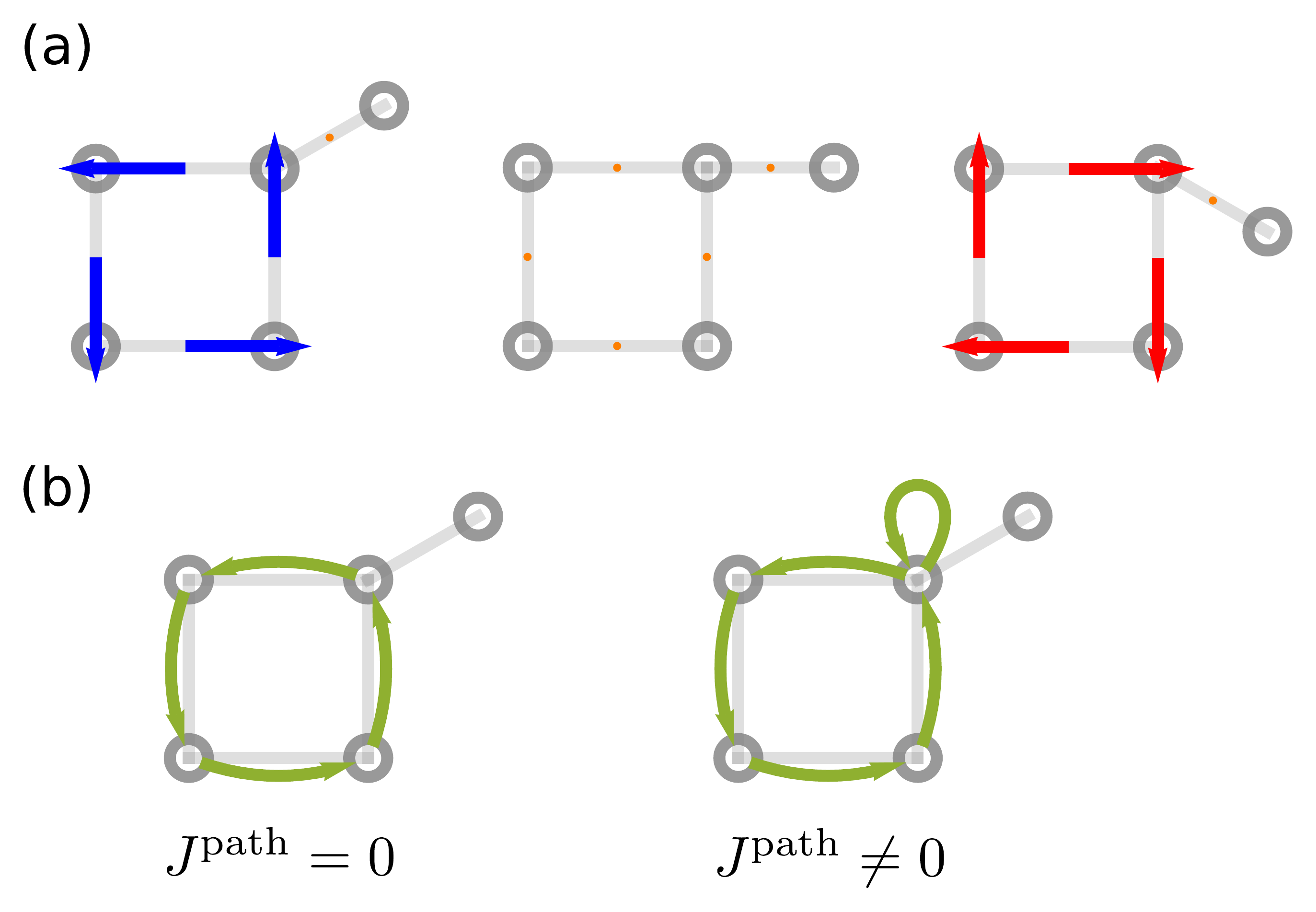}
    \caption{
        Higher-order paths in a tailed-square network.
        (a) Direction of flux can be controlled by the orientation of the side-chain.
        (b) From diagrammatic approach, the flux of the lowest-order path (square) vanishes, and the first non-vanishing path is affected by the side-chain.
    }
    \label{fig:path_square_tail}
\end{figure}

\begin{figure}[tbp]
	\centering
	\includegraphics[width=0.6\textwidth]{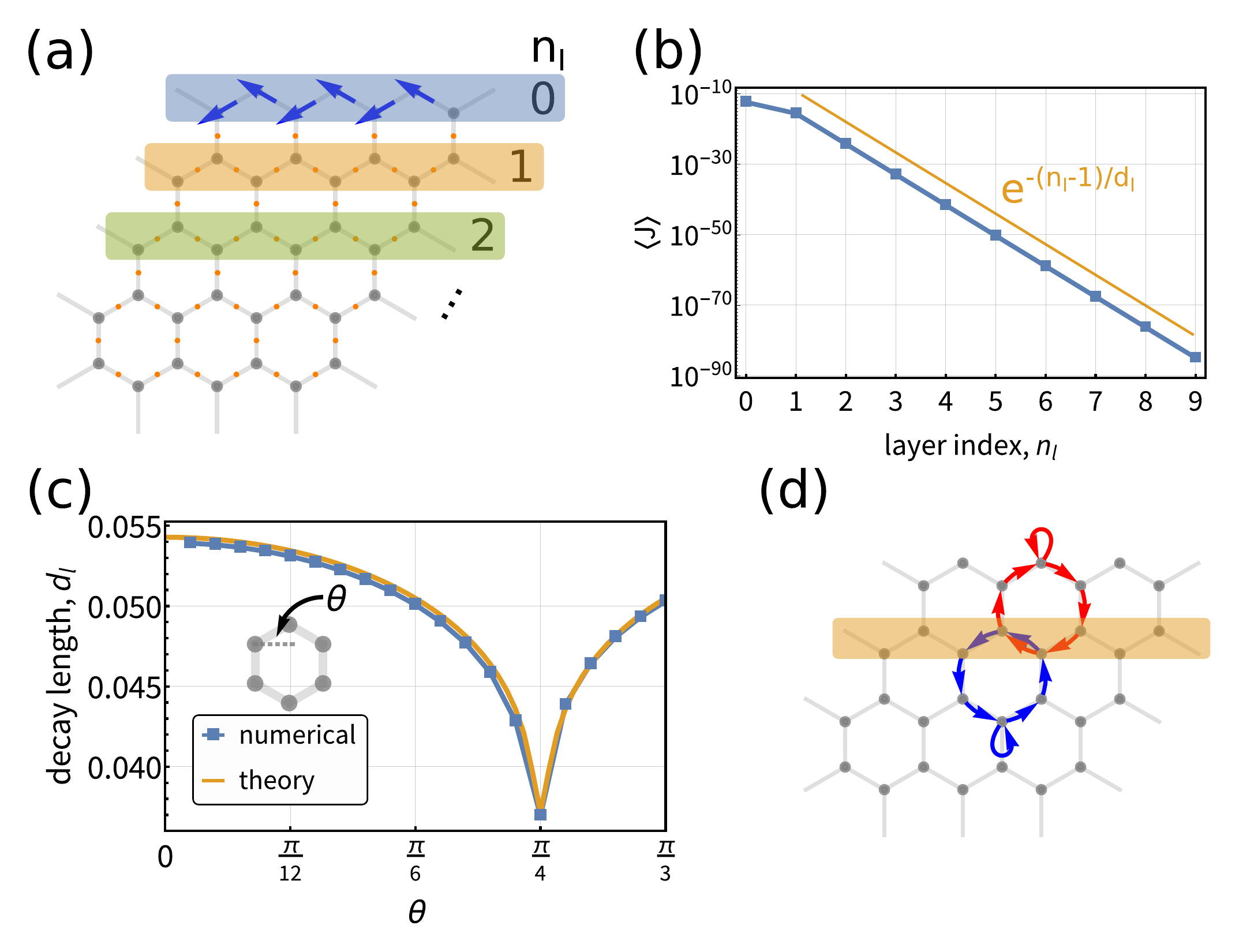}
    \caption{
        Decay of fluxes away from the boundary of honeycomb networks, and explanation using the diagrammatic technique.
        (a) Schematic of a honeycomb network which is periodic in the $x$ direction. Layers from the boundary are indexed as $n_l$.
        (b) Semi-log plot of flux $\expval{J}$ at layer $n_l$. The flux starting from layer $n_l=1$ shows exponential decay, with decay length $d_l$. Parameters used for numerical calculations are $\theta = \pi/6, k/k_0 = 0.01, \alpha = \pi/4$.
        (c) Decay length $d_l$ changes with the network angle $\theta$ non-monotonically, and the curve has a cusp at $\theta = \alpha = \pi/4$. At small $k/k_0$, perturbation theory results agree with numerical calculations.
        (d) The first non-vanishing path pair for $n_l=1$ has length $7$. The two paths do not cancel, because the loop in the bulk and at the boundary have different values.
    }
    \label{fig:path_decay}
\end{figure}

In some situations, the contribution of polygonal paths vanish, and higher order paths with loops become dominant. Unlike in polygonal paths, paths with loops are affected by side-chains.

One situation is when the polygon path itself vanishes. In \figurename~\ref{fig:path_square_tail}a, the flux of lowest order path, square (\figurename~\ref{fig:path_square_tail}b), is zero, so the main contribution comes from the path with length $5$ (\figurename~\ref{fig:path_square_tail}b). Through the loop in this path, the orientation of the side-chain controls the flux direction in the main square, without changing the geometry of the main cycle (as seen in \figurename~1b).

Another situation is that two polygon paths cancel each other, which happens in honeycomb-like networks away from the boundary (\figurename~\ref{fig:path_decay}a).
With careful calculations, the fluxes for $n_l\ge 1$ are not zero, they rather appear as an exponential decay (\figurename~\ref{fig:path_decay}b). By changing the geometric angle $\theta$, the decay length varies non-monotonically, and has a cusp at $\theta=\alpha$ (\figurename~\ref{fig:path_decay}c).
This decay and its relationship with $\theta$ can be explained by considering the paths.
While the hexagon path constitutes the lowest-order path at the boundary, it vanishes for $n_l\ge 1$ due to cancellations. The first non-vanishing pair of paths for $n_l=1$ is shown in \figurename~\ref{fig:path_decay}d, in which the loop exploits the asymmetry between the bulk side (with a vertical bond at the blue loop) and the boundary side (with no vertical bonds at the red loop). For every increment of one layer, the length of paths increases by $4$. So the flux at layer $n_l$ is on the order of $k^{4n_l+3}$, which exhibits an exponential decay $e^{-(n_l-1)/d_l}$.
Through the calculation of these paths, we get the decay length $d_l = -1/\log[4(k/k_0)^4(\sin(\theta+\alpha)\sin(\theta-\alpha))^2]$. From this result, we see that the cusp at $\theta=\alpha$ in \figurename~\ref{fig:path_decay}c is due to the term $\sin(\theta-\alpha)$. In fact, at the special point $\theta=\alpha$, paths like \figurename~\ref{fig:path_decay}d vanish, and we need to consider even higher-order paths.

\section{Simulation of active gyroscopic network coupled with a passive segment}
A simulation is shown in the Supplemental Video, which presents both the motion of particles and the energy flux through the color-labelled bonds.
The energy fluxes are in general random. During the period when $J$ is large, $J$ shows successive peaks, indicating a large energy flow from left to right. The spacing between the peaks matches the sound speed of the elastic chain ($\sqrt{k/m}$). Although the averaged direction of energy flux is from left to right, the instantaneous flux can also transport from right to left, shown as negative peaks. 

The simulation is performed using LAMMPS \cite{Plimpton1995FastParallel} with Moltemplate toolkit \cite{Jewett2013MoltemplateCoarseGrained} and custom code.
We used a Trotter splitting method \cite{Tuckerman1992ReversibleMultiple,Bussi2007AccurateSampling} to simulate the underdamped Langevin dynamics.
The integrator combines the integrator for colored noise \cite{Ceriotti2010ColoredNoiseThermostats} and that for Lorentz force \cite{Chin2008SymplecticEnergyconserving}.
We did not simulate the commonly-used overdamped Langevin dynamics, because some intricacy arises when the system also experiences a Lorentz force \cite{Chun2018EmergenceNonwhite}.
Below, we first define each step in the integrator, then present the combined result.

The velocity-Verlet step $U_{vv}$ is the integrator when both Lorentz force and the colored noise are absent. It is defined as
\begin{align}
U_{vv}(\Delta t):\quad
&v \leftarrow v + F(x) \Delta t / (2m) \\
&x \leftarrow x + v \Delta t \\
&v \leftarrow v + F(x) \Delta t / (2m),
\end{align}
where $F(x)$ is the conservative force, including on-site and inter-particle potentials.

Writing the Lorentz force part as
\begin{equation}
\dot{v} = -\pmqty{ 0 & B/m \\ -B/m & 0 } \pmqty{ v_x \\ v_y }
\equiv -a_p v ,
\end{equation}
then its integrator $U_L$ is a rotation of the velocity
\begin{equation}
    U_{L}(\Delta t):\quad
    v \leftarrow e^{-\Delta t a_p} v .
\end{equation}

Writing the colored noise part as
\begin{gather}
    \frac{d}{dt} \pmqty{ v \\ \eta }
    = -A_p \pmqty{ v \\ \eta } + B_p \pmqty{ \xi_w \\ \xi_a }, \\
    A_p = \pmqty{ \frac{\gamma}{m} & -\frac{1}{m} \\ 0 & \frac{1}{\tau} },\quad
    B_p = \pmqty{ 0 & 0 \\ 0 & \frac{\sqrt{2\gamma T_a}}{\tau} },
\end{gather}
then its integrator $U_{OUP}$ reads
\begin{equation}
    U_{OUP}(\Delta t):\quad
    \pmqty{ v \\ \eta } \leftarrow T(\Delta t) \pmqty{ v \\ \eta } + S(\Delta t) \pmqty{ 0 \\ N_a },
\end{equation}
where $N_a$ is the standard Gaussian random variable, and
\begin{align}
T(\Delta t) &= e^{-\Delta t A_p} ,\\
S(\Delta t)S(\Delta t)^T &= C_p - T(\Delta t) C_p T(\Delta t) ^T .
\end{align}
$C_p$ is the solution of $A_p C_p + C_p A_p^T = B_pB_p^T$.
$S(\Delta t)$ can be solved as an upper-triangle matrix.

Combining these steps together, the integrator for one time step $\Delta t$ reads
\begin{equation}
    U(\Delta t) = U_{OUP}(\frac{\Delta t}{2})U_{L}(\frac{\Delta t}{2})U_{vv}(\Delta t)U_{L}(\frac{\Delta t}{2})U_{OUP}(\frac{\Delta t}{2}),
\end{equation}
where the order of operations is right-to-left.

\section{Relationship between swimmer's speed and energy flux}
To understand the proportionality between $V_s$ and $\expval{J}$, we turn to the diagrammatic technique. Different from previous cases, this path sum can be computed exactly, so the result holds beyond small $k$ regime.

First we rewrite $V_s$ as the following
\begin{equation}
    \frac{V_s}{7a/24L^2} = \expval{J_{12}^s} + \expval{J_{23}^s} + \expval{J_{31}^s}, \label{eqnS:swimmer_Vs}
\end{equation}
where we have defined $\expval{J_{ij}^s} \equiv \expval{(x_i-x_j)(v_i+v_j)}$. $\expval{J_{ij}^s}$ is proportional to the energy flux via $\expval{J_{ij}} = \frac{k_{ij}}{2}\expval{J_{ij}^s}$, where $k_{12}=k_{23}=k$, and $k_{31}=0$ (because $\expval{J_{31}} = 0$, there is no energy flux from $3$ to $1$).
We see that both $\expval{J_{12}^s}$ and $\expval{J_{23}^s}$ are proportional to the flux $\expval{J}$ apart from a factor $k$, so the remaining task is to find the relationship between $\expval{J_{31}^s}$ and $\expval{J}$ or $\expval{J_{12}^s}$.

\begin{figure}[tbp]
	\centering
	\includegraphics[width=0.4\textwidth]{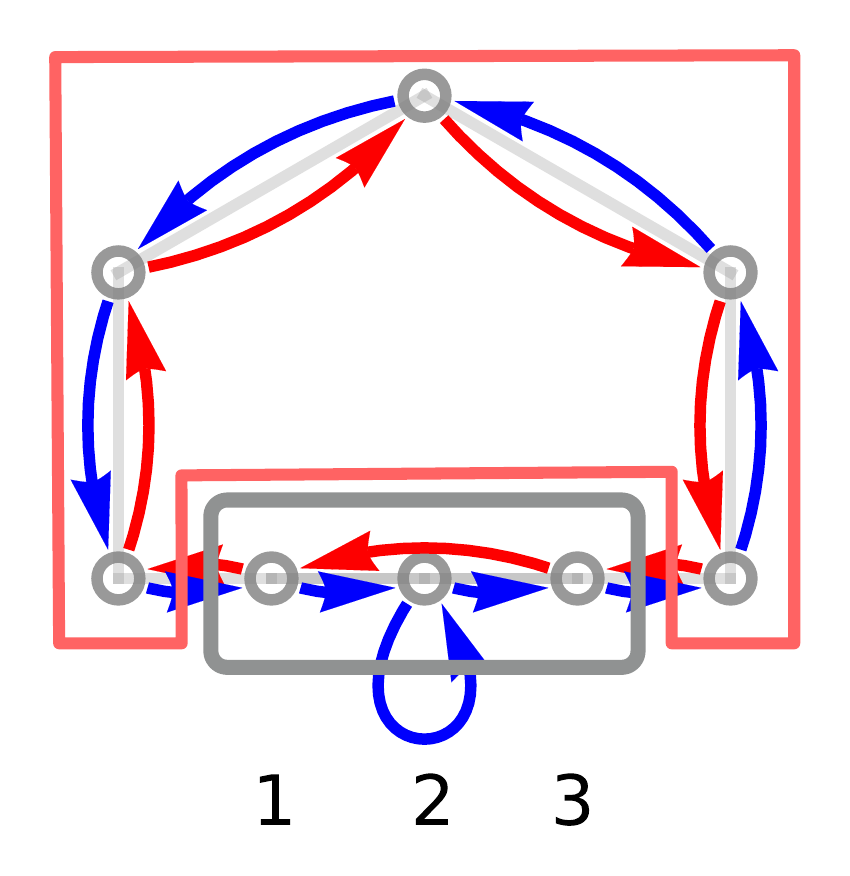}
    \caption{
        One example of $J_{31}^s$ path (red) and $J_{12}^s$ path (blue). Passive particles are boxed in gray, and the active ones are boxed in red.
    }
    \label{fig:swimmer_path}
\end{figure}

We use a diagrammatic technique with the modification that the paths should contain only one $3\rightarrow 1$ segment. This modification is a consequence of the fact that particle $3$ and $1$ are not bonded.
We now illustrate the correspondence between the paths for $\expval{J_{31}^s}$ and for $\expval{J_{12}^s}$. For each path $l$ for $\expval{J_{31}^s}$, we can construct $n$ paths for $\expval{J_{12}^s}$ by reversing $l$ then replacing $1\rightarrow 3$ by $1\rightarrow 2(\rightarrow 2)^n \rightarrow 3$, where $n=0,1,\dots$. An example construction of paths is shown in \figurename~\ref{fig:swimmer_path}.
For $\expval{J_{12}^s}$, all its paths can be constructed in this way.
As a result, there is a $1$ to $n$ correspondence between the paths for $\expval{J_{31}^s}$ and for $\expval{J_{12}^s}$, which leads to the relationship
\begin{equation}
    \expval{J_{12}^s} = \frac{k}{k_0}\sum_{n=0}^\infty (-2\frac{k}{k_0})^n (-\expval{J_{31}^s}) = \frac{k/k_0}{1-(-2k/k_0)} (-\expval{J_{31}^s}), \label{eqnS:swimmer_path_mapping}
\end{equation}
where $k_0 = k_g + m/\tau^2$ ($B,\gamma=0$ for the passive part), and the factor $-2\frac{k}{k_0}$ comes from the loop $2\rightarrow 2$.
Plugging \eqnname~\eqref{eqnS:swimmer_path_mapping} to the expression of $V_s$ \eqnname~\eqref{eqnS:swimmer_Vs}, we obtain the proportionality
\begin{equation}
    \frac{V_s}{7a/24L^2} = -\frac{k_0}{k} \frac{\expval{J}}{k/2},
\end{equation}
which is \eqnname~(15) in the main text.
Since we have considered all the paths, this result can be analytically continued to arbitrarily large $k$.

From this diagrammatic technique we also see that, the proportionality constant is independent of the geometry of the active part of the network. This is because the paths through the active part for $\expval{J_{31}^s}$ and for $\expval{J_{12}^s}$ are identical.

%